\documentclass[10pt,aps,prl,twocolumn,superscriptaddress,amsmath,amssymb,shortbibliography]{revtex4-2}
\usepackage{graphicx}
\usepackage{color}
\usepackage{xcolor}
\usepackage{comment}
\usepackage{mathtools}

\usepackage[colorlinks=false, linkbordercolor=red, citebordercolor=green, urlbordercolor=cyan, pdfborderstyle={/S/U/W 1}]{hyperref}

\usepackage{soul}

\begin{document}

\title{Quantum non-Gaussianity of multi-phonon states of a single atom}

\author{L.~Podhora}
\affiliation{Department of Optics, Palack\'y University, 17. listopadu 12, 77146 Olomouc, Czech Republic}
\author{L.~Lachman}
\affiliation{Department of Optics, Palack\'y University, 17. listopadu 12, 77146 Olomouc, Czech Republic}
\author{T.~Pham}
\affiliation{Institute of Scientific Instruments of the Czech Academy of Sciences, Kr\'{a}lovopolsk\'{a} 147, 612 64 Brno, Czech Republic}
\author{A.~Le\v sund\'ak}
\affiliation{Institute of Scientific Instruments of the Czech Academy of Sciences, Kr\'{a}lovopolsk\'{a} 147, 612 64 Brno, Czech Republic}
\author{O.~\v C\'ip}
\affiliation{Institute of Scientific Instruments of the Czech Academy of Sciences, Kr\'{a}lovopolsk\'{a} 147, 612 64 Brno, Czech Republic}
\author{L.~Slodi\v{c}ka}
\email{slodicka@optics.upol.cz}
\affiliation{Department of Optics, Palack\'y University, 17. listopadu 12, 77146 Olomouc, Czech Republic}
\author{R.~Filip}
\email{filip@optics.upol.cz}
\affiliation{Department of Optics, Palack\'y University, 17. listopadu 12, 77146 Olomouc, Czech Republic}

\date{\today}

\begin{abstract}
Quantum non-Gaussian mechanical states are already required in a range of applications. The discrete building blocks of such states are the energy eigenstates - Fock states. Despite progress in their preparation, the remaining imperfections can still invisibly cause loss of the aspects critical for their applications. We derive and apply the most challenging hierarchy of quantum non-Gaussian criteria on the characterization of single trapped-ion oscillator mechanical Fock states with up to 10~phonons. We analyze the depth of these quantum non-Gaussian features under intrinsic mechanical heating and predict their requirement for reaching quantum advantage in the sensing of a mechanical force.
\end{abstract}


\maketitle

\section{Introduction}

Long-term progress in nonlinear control of quantum mechanical states allows the deterministic generation of discrete quanta of energy embedded in a single motional mode~\cite{leibfried1996experimental,meekhof1996generation,roos1999quantum,toyoda2015hong,kienzler2017quantum, ding2017quantum,zhang2018noon,fluhmann2019encoding,mccormick2019quantum,chu2018creation}. They find direct application in quantum sensing~\cite{gessner2019metrological,wolf2019motional}. This effort has opened many ways to generate other quantum non-Gaussian (QNG) states broadly useful in quantum metrology~\cite{hanamura2021estimation,mccormick2019quantum,duivenvoorden2017single} and quantum error correction~\cite{michael2016new,hu2019quantum,campagne2020quantum,fluhmann2019encoding}. QNG aspects have also appeared as a consequence of nonlinear dynamics on oscillator ground states as well as in the broadly discussed quantum engines~\cite{ghosh2018thermodynamic, maslennikov2019quantum} and recently, in quantum phase transitions~\cite{cai2021observation}. Nonlinearity is always required in some form, since QNG properties can never arise from any mixture of Gaussian states produced by linearized dynamics~\cite{hudson1974wigner,walschaers2021non}. In a continuous-variable representation, QNG states can exhibit a spectrum of negative values of a Wigner function~\cite{schleich2011quantum}. For optical tests of QNG features belonging to Fock states with an unambiguous identification the possibilities beyond a continuous-variable representations based on the Wigner function, hierarchies of criteria for photonic states have been derived and experimentally verified for up to three photons~\cite{lachman2019faithful}. Recently, the Husimi function has also been used to define a stellar representation of QNG states~\cite{chabaud2020stellar}. They specifically accent the optical loss being the main limitation of propagating states of light. Note that the condition for genuine n-photon quantum non-Gaussianity introduced in~\cite{lachman2019faithful} is equivalent to the stellar rank for Fock states~\cite{chabaud2021certification}.

Here, we experimentally demonstrate the most strict hierarchy of QNG criteria suitable for individual Fock states of mechanical systems, where mechanical heating is the critical limitation~\cite{brownnutt2015ion, aspelmeyer2014cavity}. Therefore, Fock states of mechanical oscillators crucially deserve different criteria and analysis than states of light~\cite{higginbottom2016pure,lachman2019faithful,straka2018quantum,straka2014quantum,ra2020non,walschaers2018tailoring,chabaud2021certification}. The $n$-th order criteria applied to the phonon-number distribution can conclusively recognize that analyzed aspect of the oscillator state is not reachable by any mixture of $D(\alpha)S(r) \sum_{m<n-1}^{} c_{m} |m\rangle$ with complex $\alpha$, $r$ and $c_m$, where $D(\alpha)$ and $S(r)$ are Gaussian displacement and squeezing operations applied to any superposition of Fock states up to~\textit{n}-1, and $\sum_{m}^{} |c_{m}|^2 = 1$. It proves that a genuine Fock state $|n\rangle$ has been unambiguously generated and therefore the observed state can supply applications powered by its exclusive QNG aspects.

We verify this hierarchy~\cite{lachman2019faithful,chabaud2021certification} on states approaching Fock states for a motional mode of a trapped ion oscillator and provide a reference methodology for such an implementation together with analysis of the effect of the dominant deteriorating mechanism in mechanical systems - thermal heating. The resulting unprecedentedly high genuine QNG properties and the paramount possibility of their verification allow for an unambiguous analysis of the states powering applications requiring genuine QNG states of atomic motion for quantum enhanced sensing~\cite{wolf2019motional,mccormick2019quantum} or assist prospective directions of efficient qubit encoding and error correction~\cite{fluhmann2019encoding,de2022error}, which require a clear and simple threshold-based experimental methodology. Although these works provided pioneering proof-of-principle demonstrations of sensing applications which provably require high probability of the Fock states~\cite{wolf2019motional,mccormick2019quantum}, the feasibility, robustness, and experimental relationship to applications of the genuine QNG aspects could not be directly accessed or proven, despite their crucial role.
We generate motional states of a single trapped ion with up to ten motional quanta to develop and test the methodology for the conclusive and unambiguous analysis of their genuine QNG properties~\cite{lachman2019faithful,chabaud2021certification}. We estimate their robustness to the heating relevant to mechanical systems and discover that the $n$-th order criteria are more strict than basic quantum non-Gaussianity~\cite{filip2011detecting}, negativity of the Wigner function and even observation of the largest negative annulus in the Wigner function corresponding to the Fock state~$|n\rangle$~\cite{zapletal2021experimental}. 
We confirm that QNG features correspond to the crucial and challenging resource powering quantum sensing of coherent displacement or recoil heating~\cite{wolf2019motional} by the estimation of sensing capability for the generated states and realize that genuine QNG features are indeed necessary for reaching high sensitivity.

\begin{figure*}[!ht]
\begin{center}
\includegraphics[width=1.5\columnwidth]{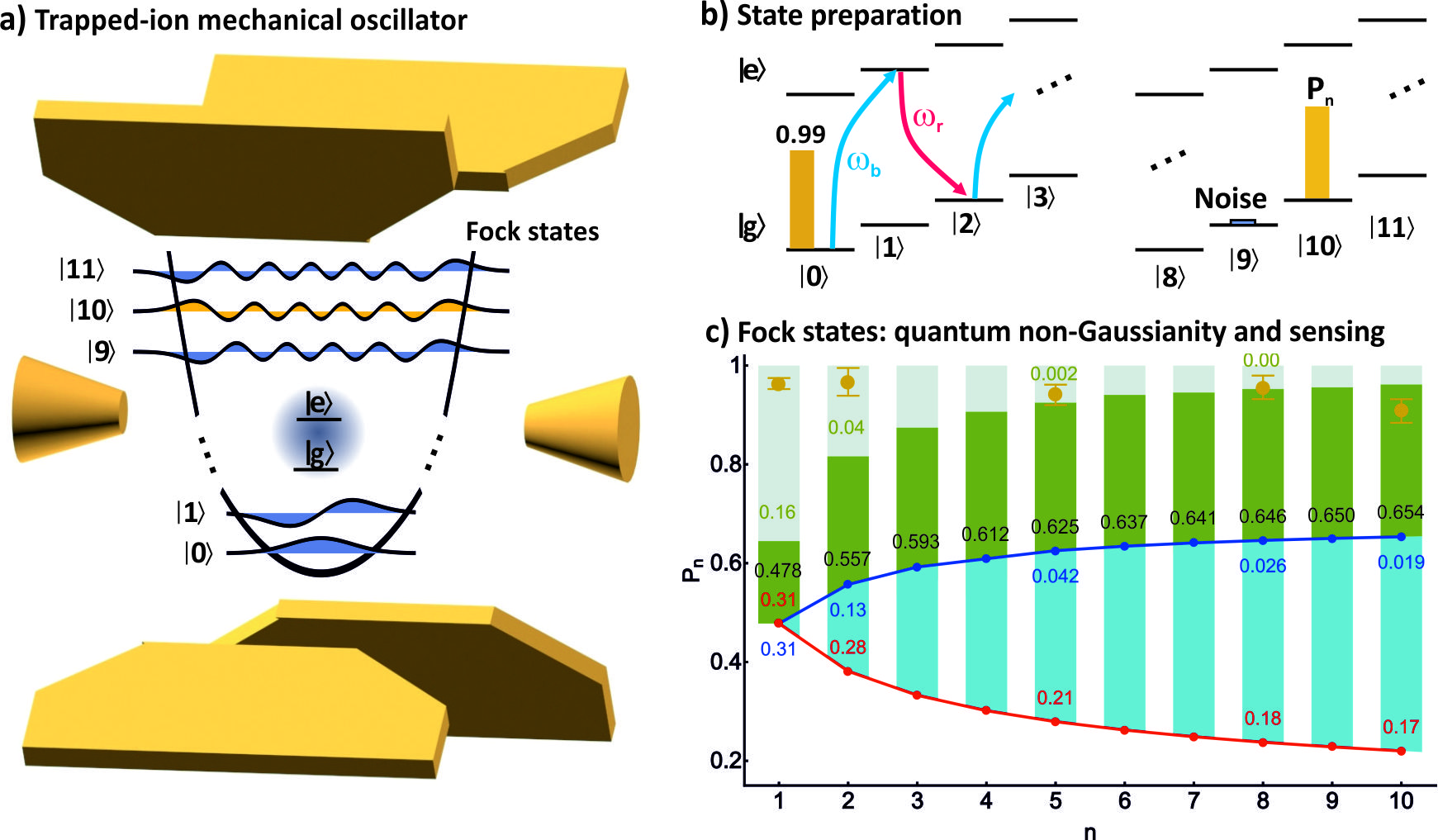}
\caption{
a) The mechanical oscillator corresponds to axial harmonic motion of a single $^{40}$Ca$^+$ ion localized in a linear Paul trap. The generation and analysis of states approaching idealized Fock states illustrated by their corresponding wave functions is implemented through interaction of the electronic ground $|{\rm g}\rangle$ and metastable $|{\rm e}\rangle$ states with the quantized harmonic motion on the first motional sidebands. b)~The sequence for preparation of genuine QNG states includes initialization to the electronic and motional ground state $|{\rm g},0\rangle$ followed by deterministic transfer of the accumulated population to number states by repetitive coherent excitation of the blue and red sidebands depicted as blue and red arrows, respectively. The spectroscopic analysis of phonon number populations $P_n$ around the target Fock state implements an unambiguous identification of the genuine QNG features. c)~Characterization of the Fock states of mechanical motion. The yellow points represent the measured populations $P_{n}$ for the experimentally generated states. The thresholds for genuine $n$-phonon QNG are represented by blue points with the corresponding black numbers showing their numerical value~\cite{lachman2019faithful,chabaud2021certification}. The associated blue numbers quantify the thermal depth of genuine $n$-phonon QNG. 
Similarly, the red points identify thresholds for observation of basic QNG aspects~\cite{filip2011detecting} and the associated red numbers determine their thermal depth. The green bars depict the force estimation capability of a specific model of noisy Fock states, where the probability $P_n$ exceeding the presented threshold values certifies a metrological advantage~\cite{wolf2019motional} against the previous ideal Fock state $|n-1\rangle$, while the corresponding numbers quantify the thermal depth of this advantage for the measured states.}
\label{fig:setup}
\end{center}
\end{figure*}

\section{Hierarchy of QNG criteria for mechanical Fock states }


Genuine $n$-phonon QNG defines phonon statistics that do not evolve from Gaussian processes modifying any superposition involving Fock states lower than $|n\rangle$. It determines a quantum aspect that these statistics share with the state~$|n\rangle$, which cannot be achieved with all the lower Fock states. Formally, the genuine $n$-phonon QNG of pure states requires
\begin{equation}
|\psi_{n}\rangle \neq D(\alpha) S(r) \sum_{m=0}^{n-1} c_{m} |m\rangle,
\label{eq:nonGaussian}
\end{equation}
where $S(r)=\exp\left[r \left(a^{\dagger}\right)^2-r^* a^2\right]$ and $D(\alpha)=\exp\left[\alpha^* a-\alpha a^{\dagger}\right]$ are squeezing and  displacement unitary operators, respectively, with a complex squeezing parameter $r$, complex displacement $\alpha$, and $a,a^{\dagger}$ corresponding to ladder operators~\cite{lachman2019faithful}. Both $D(\alpha)$ and $S(r)$ are Gaussian operations which correspond to linear transformations of the ladder operators and don't increase the non-Gaussianity of a state~$\sum_{m=0}^{n-1}c_m|m\rangle$. An impure state possesses this quantum aspect when it does not correspond to any statistical mixture of the right side of~(\ref{eq:nonGaussian}). The most strict criterion uses a bosonic distribution $P_{n} = \langle n |\rho| n \rangle$ to recognize the genuine $n$-phonon QNG for the imperfect Fock state $|n\rangle$ when the probability $p_{n}$ exceeds the threshold
\begin{equation}
\overline{p}_{n} = \max_{\alpha,r,c_0,...,c_m} |\langle n|D(\alpha) S(r) \sum_{m=0}^{n-1} c_{m} |m\rangle|^{2}.
\label{eq:threshold}
\end{equation}
The linearity of the maximizing task for determining $\overline{p}_{n}$ guarantees that any mixture of rejected states cannot surpass $\overline{p}_{n}$. More details about the genuine QNG criteria suitable for mechanical systems can be found in the~Supplementary information~S1. We note that identical threshold values for $p_n$ can be alternatively derived using the stellar representation of QNG states~\cite{chabaud2021certification}, which in addition, provide analytical expressions for the thresholds for the considered Fock states.

A trapped-ion system with the possibility of unprecedented motional control and practically lossless readout is a suitable candidate for tests of the diverse QNG aspects of motion in a well-defined mode and with high experimental precision~\cite{leibfried1996experimental, roos1999quantum, toyoda2015hong, kienzler2017quantum, ding2017quantum, zhang2018noon, fluhmann2019encoding,mccormick2019quantum}. It allows for implementations with minimal added noise and promises prospects of accessing QNG features with feasible observations of the sensitive properties relevant for applications~\cite{fluhmann2019encoding,wolf2019motional,mccormick2019quantum,gan2020hybrid}.
The experiment is implemented on the axial motional mode of a single $^{40}$Ca$^+$ ion held in a linear Paul trap. Figs.~\ref{fig:setup}-a) and b) show a simplified scheme of the experimental arrangement and the experimental sequence~\cite{S2}. The quantum motional states are generated by a transfer of the initial vacuum state to the state approaching the Fock state $|n\rangle$ by an iterative sequence of motional population raising operations corresponding to blue and red sideband $\pi$-pulses on the $|{\rm g}\rangle \sim 4{\rm S}_{1/2}(m=-1/2) \leftrightarrow |{\rm e}\rangle\sim 3{\rm D}_{5/2}(m=-1/2)$ electronic transition with frequencies $\omega_b$ and $\omega_r$, respectively~\cite{leibfried1996experimental,roos1999quantum,mccormick2019quantum}. The state preparation is followed by estimation of the phonon-number probabilities $P_{n}$ using the precise analysis of Rabi oscillations on the first blue motional sideband~\cite{leibfried2003quantum}.

Fig.~\ref{fig:setup}-c) analyses the exhibition of genuine $n$-phonon quantum non-Gaussianity using idealized and measured Fock states (yellow data points).
The simulations signify that the observability of the lowest QNG features (red points) characterized by $|\psi_{n}\rangle \neq D(\alpha) S(r) |0\rangle$~\cite{filip2011detecting} in the presence of heating monotonously increases with $n$ for ideal Fock states. On the contrary, the generation and observation of genuine QNG (blue points) for high Fock states inside the hierarchy~(\ref{eq:nonGaussian},\ref{eq:threshold}) is challenging and its sensitivity to imperfections in the state preparation and detection increases with $n$. The ideal thermalization dynamics considered, corresponding to a Gaussian additive noise, can be broadly employed for an estimation of the QNG depth, analogous to how the damping was used for photonic implementations~\cite{straka2014quantum}. The thermal depth has been evaluated as the corresponding increase of the mean thermal energy $\bar{n}_{\rm th}$ for the same thermalization strength applied to the vacuum state. This allows for an platform-independent comparison of the genuine QNG states in mechanical systems. In turn, these measurements can also be employed for testing the quality of the mechanical system or for sensing the amount of inherent thermal noise. The thermal depth of the measured states is much smaller for the genuine QNG hierarchy~(\ref{eq:nonGaussian},\ref{eq:threshold}), in contrast to the lowest QNG criteria~\cite{filip2011detecting} which are actually less demanding for higher Fock states. Still, the measured 10-phonon states conclusively proved the genuine QNG features. Although the absolute thermal depths shown as mean phonon numbers $\bar{n}_{\rm th}$ decrease both for the genuine QNG features (blue values) and for the lowest QNG criteria~\cite{filip2011detecting} (red values), their ratio increases to about an order of magnitude for $n=10$. A further analysis 
can be found in the Supplementary information~S3 and~S4.

\section{Robustness of the genuine QNG}

In the vast majority of implementations of quantum mechanical oscillators with atoms and electromechanical systems, their manipulation and observation is accompanied by small heating~\cite{brownnutt2015ion, aspelmeyer2014cavity,chu2018creation}. The heating mechanisms relevant for the single-ion oscillator include photon recoil due to the spontaneous emission or interaction of the charged particle with the surrounding thermal trapping environment. 
On the short time scales relevant for the generation of target Fock states, the heating results in the error given predominantly by $P_{n-1}+P_{n+1}$. This assumption is confirmed by our observations of motional populations after the state preparation presented in the Supplementary information~S2. 
The depth of QNG for states of mechanical systems can be naturally defined as the amount of heating necessary to reach the corresponding QNG boundary~(\ref{eq:nonGaussian},\ref{eq:threshold}).

\begin{figure*}[!th]
\begin{center}
\includegraphics[width=2.1\columnwidth]{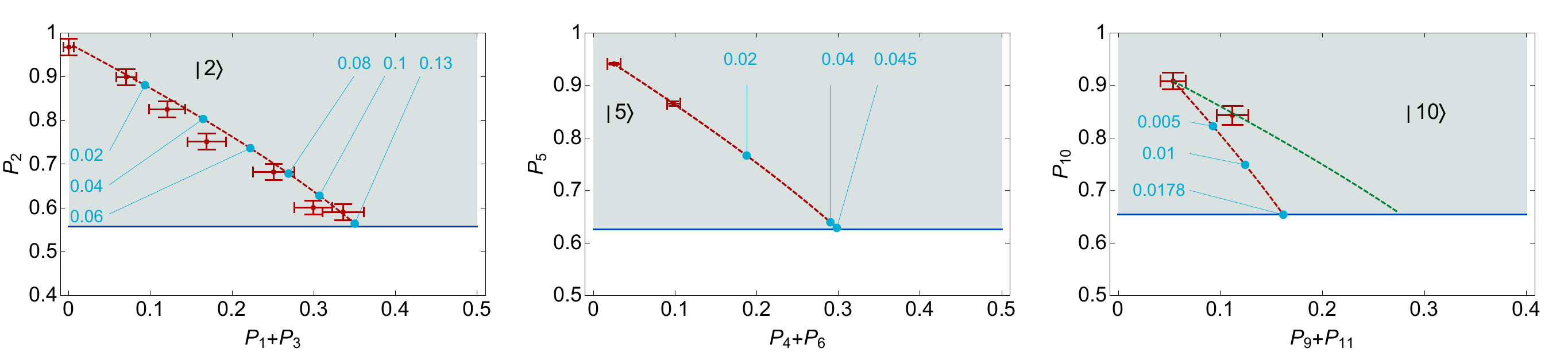}
\caption{Measurements of the QNG depth for states approaching Fock states $|2\rangle$, $|5\rangle$, and $|10\rangle$. Each graph shows the probability $P_{n}$ and the overall probability $P_{n-1}+P_{n+1}$ relevant for the evaluation of the thermalization. The red points represent the measured states after several heating steps with an increasing length of the 397~nm laser pulse. The red dashed lines correspond to the theoretical model of ion heating by random photon recoils, 
with blue points giving a scale of the applied thermalization using an equivalent mean thermal phonon number $\bar{n}_{\rm th}$ when applied to the vacuum state. The green dashed line in a graph for state approaching $|10\rangle$ shows the simulated trajectory above the Doppler cooling limit and horizontal \textbf{blue} lines depict thresholds of the $n$-th order genuine QNG~\cite{lachman2019faithful,chabaud2021certification} given by~(\ref{eq:nonGaussian},\ref{eq:threshold}). For $n=2$ and 5 the green dashed lines coincide with red dashed one.}
\label{fig:QNG_robustness}
\end{center}
\end{figure*}

We study the QNG depth by the evaluation of the robustness of the generated QNG states to the mechanical heating induced by random photon recoils 
implemented by illuminating the ion with a short laser pulse on the 4S$_{1/2} \leftrightarrow {\rm 4P}_{1/2}$ transition. The heating rate has been set to $(115\pm 2)\times10^3$~ph/s on an ion prepared in the motional ground state. Fig.~\ref{fig:QNG_robustness} shows measurements of the relative depth scaled consistently to the mean added thermal energy when the same thermalization is applied to the vacuum state. The simulations employ a single-parameter photon recoil model, shown as red dashed curves. However, as the employed laser beam is set to optimize the Doppler cooling, the trajectories in the $P_{n}$,$P_{n-1}+P_{n+1}$ space are expected to deviate from this model for high initial Fock states or for very long thermalization times. A finite temperature reservoir characterized by two parameters~\cite{eschner2003laser} has to be included to plausibly explain the evolutions in these limits (green dashed curve). 
The observations confirm the predicted increased sensitivity of states with high order $n$ of genuine QNG to heating, as can be seen by the gradually reduced scale of the effective mean thermal phonon number $n_{\rm th}$. A detailed description of thermalization can be found in the Supplementary information~S3.

\section{Force estimation capability}

\begin{figure}[!th]
\begin{center}
\includegraphics[width=1.\columnwidth]{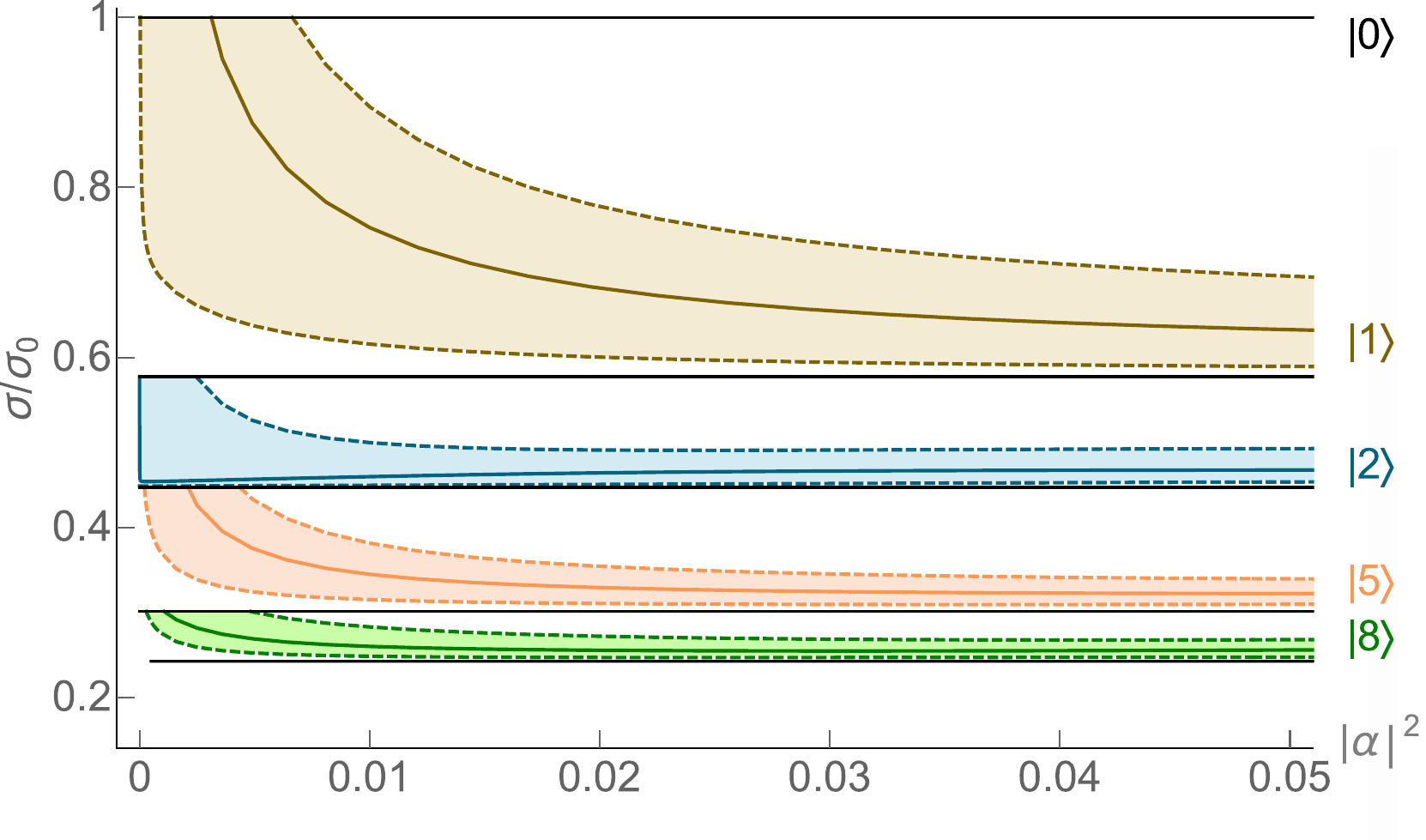}
\caption{Estimation of metrological advantage of realized states for sensing of a small force. The horizontal axis quantifies the amplitude in the phase space that the force causes. The vertical axis shows the minimal standard deviation $\sigma$ estimated by optimization of the Fisher information in the Eq.~\ref{eq:fisher} normalized to $\sigma_0$ resulting from sensing using a motional ground state. The black lines show $\sigma/\sigma_0$ for ideal Fock states. The brown, blue, orange, and green solid curves correspond to prepared states approaching the Fock states $|n\rangle=|1\rangle, |2\rangle$, $|5\rangle$, and $|8\rangle$, respectively.  
The colored regions show $\sigma/\sigma_0$ for states with the phonon-number distributions within experimental error bars.}
\label{fig:forceEstimation}
\end{center}
\end{figure}

A mechanical oscillator in a state approaching Fock state can be directly used for a phase-insensitive sensing of a weak force causing a tiny displacement $\alpha$~\cite{wolf2019motional}. Let the oscillator be prepared in an initial state $\rho$ and let $D(\alpha)$ denote the displacement operator characterizing the evolution that the force induces. The Fisher information for the estimation of the parameter $\vert \alpha \vert^2$ reads
\begin{equation}
F=\sum_{m=0}^{\infty}\frac{1}{P_m(\vert \alpha \vert^2)}\left[\frac{\mathrm d}{{\mathrm d}_{\vert \alpha \vert^2}} P_m(\vert \alpha \vert^2)\right]^2,
\label{eq:fisher}
\end{equation}
where $ P_n(\vert \alpha \vert^2) =\langle n \vert D(\vert \alpha \vert) \rho D^{\dagger}(\vert \alpha \vert)  \vert n \rangle $ is the phonon-number distribution on which the sensing is performed. 
We evaluate the $\sigma$ saturating the Cram\'er-Rao bound for the realized states. 
The metrological advantage $R_{\rho}$ can be quantified according to
\begin{equation}
R_{\rho}(\vert \alpha \vert^2)=\frac{\sigma}{\sigma_0},
\label{eq:metrolAdv}
\end{equation}
where $\sigma_0$ stands for the standard deviation of the measurement for the mechanical probe prepared in the motional ground state. For ideal Fock states, (\ref{eq:metrolAdv}) approaches
\begin{equation}
R_{\vert n \rangle}(\vert \alpha \vert^2)=\frac{1}{\sqrt{2n+1}},
\end{equation}
which is independent of the estimated $\vert \alpha \vert^2$. The full derivation can be found in the Supplementary information~S5. 
If a state $\rho$ achieves $R_{\rho}(\vert \alpha \vert^2)<R_{\vert n \rangle}$ for some $\vert \alpha \vert^2$, it possesses a capability to surpass the sensing with the Fock state $\vert n \rangle$. Therefore, the sequence $R_{\vert n \rangle}$ establishes a hierarchy of conditions for sensing classifying the states approaching Fock states. Fig.~\ref{fig:forceEstimation} presents the metrological potential of realized states to pass these conditions. Specifically, realistic Fock states up to $n=10$ surpass the limit given by the vacuum state and the prepared state approaching the ideal Fock state $|8\rangle$ presents the capacity to exceed the Fock state $|5\rangle$. The realized state approaching $|10\rangle$ and all the higher prepared Fock states did not possess the metrological advantage against the ideal Fock state $|8\rangle$ due to noise mainly including the residual heating during the state preparation, see Supplementary information~S2.

For sensing of a small $\vert \alpha \vert^2$, the noise affects how far a prepared state is from the threshold given by the ideal Fock state $|n\rangle$. The advantage gets lost even for a very small noise in the high $|n\rangle$ limit. $R_{\rho}(|\alpha|^2)$ tends to saturate for displacement on the order of $10^{-2}$ and approaches the gain expectable for ideal Fock states. At the same time, employment of realistic states with high $n$ in the limit of small displacements seems to be further favoured due to effectively decreasing dependence of the offset in $R_{\rho}(\vert \alpha \vert^2)$ on $n$, when compared to ideal Fock states. This has been confirmed by a numerical simulation considering sensing with states resulting from thermalization of Fock states, 
see the Supplementary information S5.
We note the visible deviation from this trend for the state approaching $|2\rangle$ caused by the analysis on the phonon distribution with residual $P_0~\sim 1\,\% $, instead of the mixture of $P_1$ and $P_3$ expectable for the case of dominant spread through thermalization. However, these small populations are of a low statistical significance.

\section{Conclusions}

The presented demonstration of high-order genuine QNG states~\cite{lachman2019faithful,chabaud2021certification} provides the first feasible methodology for conclusive, hierarchical, and sensitive evaluation of mechanical Fock states. Simultaneously, it predicts development of methods for quantum-enhanced sensing of motional heating or coherent displacements on quantum mechanical oscillators~\cite{wolf2019motional}. Beyond these direct applications, it provides a fundamental milestone of experimental witnessing of intrinsic properties of highly nonclassical states applicable to a broad spectrum of applications of quantum-enhanced control and measurement of mechanical motion in optical frequency metrology~\cite{hempel2013entanglement,wan2014precision}, quantum error correction~\cite{de2022error,michael2016new,hu2019quantum,campagne2020quantum}, tests of quantum thermodynamics~\cite{gely2019observation,gelbwaser2015work}, or gravitational wave detection~\cite{tse2019quantum}. Together with the recent demonstration of the viable preparation of states suggesting proximity to number states with up to $|n\rangle\sim 100$ in the same experimental platform~\cite{mccormick2019quantum}, the presented approach will allow for optimization and comparison of these quantum states across different sensing scenarios with a clearly identifiable fundamental and application relevance~\cite{gorecki2020optimal,kubica2021using}. The provided evaluation of provable quantum metrological advantage of the generated states for sensing of small displacements with motional probes close to ideal Fock states $|n\rangle$ for $n$ up to 8 suggests the potential of the presented methodology as a tool for analysis of the preparation of large genuine QNG sensing states with energies and corresponding metrological stability gains substantially beyond current experimental capabilities of quantum sensing with atomic and mechanical oscillators~\cite{wolf2019motional,schreppler2014optically,hebestreit2018sensing}.

\begin{acknowledgments}
L.~L., L.~S., and R.~F. are grateful to the support from the Czech Science Foundation under the project GA21-13265X. L.~P., T.~P., A.~L., and O.~C., acknowledge the
support by Grant No.~GA19-14988S from the Czech Science Foundation and the project CZ.02.1.01/0.0/0.0/16\underline{ }026/0008460 of MEYS CR. T. P., A. L. and O. C. acknowledge the project EMPIR 20FUN01 TSCAC, which received funding from the EMPIR programme co-financed by the participating States and from European Union's Horizon 2020 research and innovation programme. The research leading to these results has received funding from the H2020 European Programme under Grant Agreement No.~951737 NONGAUSS. L. P. acknowledges the internal project of Palacky University IGA-PrF-2021-006.
\end{acknowledgments}


\clearpage

\newpage

\renewcommand{\theequation}{S\arabic{equation}}
\renewcommand{\thefigure}{S\arabic{figure}}
\renewcommand{\thesection}{S\arabic{section}}
\renewcommand{\theHequation}{Supplement.\theequation}
\renewcommand{\theHfigure}{Supplement.\thefigure}
\renewcommand{\bibnumfmt}[1]{[S#1]}
\renewcommand{\citenumfont}[1]{S#1}

\setcounter{equation}{0}
\setcounter{figure}{0}
\setcounter{table}{0}
\setcounter{section}{0}
\setcounter{page}{1} \makeatletter

\onecolumngrid

\section*{Supplementary information: \\ Quantum non-Gaussianity of multi-phonon states of a single atom}

\twocolumngrid

\section{Derivation of the criteria for genuine $n$-photon quantum non-Gaussianity}
\label{SI:criteria}

Employment of the detection process capable of resolving the number of detected motional quanta allows us to identify criteria imposing a condition on the probability $P_{n}$ of detecting $n$ phonons~\cite{lachman2019faithfulS}. The genuine $n$-phonon quantum non-Gaussianity can be unambiguously recognized when $P_{n}>\bar{p}_n$ with
\begin{equation}
\bar{p}_n=\max_{\alpha,r,c_0,...,c_m} |\langle n |D(\alpha) S(r) \sum_{m=0}^{n-1} c_m |m\rangle |^2,
\label{Fa}
\end{equation}
where the maximizing is performed over acting of the squeezing operator $S(r)=\exp\left[r \left(a^{\dagger}\right)^2-r^* a^2\right]$ and the displacement operator $D(\alpha)=\exp\left[\alpha^* a-\alpha a^{\dagger}\right]$ on the core state $\sum_{m=0}^{n-1} c_m |m\rangle$. The maximum happens when $\alpha$, $r$ and $c_0,...,c_{n-1}$ become real parameters that obey
\begin{align}
   &\partial_{\alpha} \vert \langle n \vert D(\alpha)S(r) \sum_{k=0}^{n-1} c_k |k\rangle \vert^2=0 \label{alf}\\
    &\partial_{r} \vert \langle n \vert D(\alpha)S(r) \sum_{k=0}^{n-1} c_k |k\rangle \vert^2=0 \label{r}\\
    &\partial_{c_{n-1}} \vert \langle n \vert D(\alpha)S(r) \sum_{k=0}^{n-1} c_k |k\rangle \vert^2=0 \label{cn1}\\
    &\vdots \nonumber\\
    &\partial_{c_{1}} \vert \langle n \vert D(\alpha)S(r) \sum_{k=0}^{n-1} c_k |k\rangle \vert^2=0 \label{c1}
\end{align}
with the constraint $c_0^2=1-\sum_{k=0}^{n-1} c_k^2$. Since the maximizing parameters satisfy the relations $c_{n-1}> c_{n-2}> ...> c_0$, it can be achieved by consecutive solving Eqs.~(\ref{alf}-\ref{c1}) using the following steps. Firstly, an algorithm searches for $\alpha$ and $r$ satisfying (\ref{alf}) and (\ref{r}) for $c_{n-1}=1$. Further, it makes correction of the parameters $c_1,...,c_{n-1}$ by solving sequentially the equation in (\ref{cn1}) for $c_{n-1}$ and all the following equations denoted symbolically by the dots up to the equation (\ref{c1}). Finally, the whole procedure is performed once more with the updated parameters $c_1,...,c_{n-1}$. This allows us to acquire $\bar{p}_n$ with high numerical accuracy. The thresholds obtained by this method are presented in the Fig.~1 in the main part of the manuscript. We note that analogous criteria with identical threshold values can be obtained with approach based on a stellar representation of QNG states~\cite{chabaud2021certification}. In addition, they provide an analytical expressions for thresholds for considered Fock states.

\section{Experimental details}
\label{SI:experiment}

\subsection{State preparation}

The experiment is performed on the single mechanical oscillator implemented using a well controllable motion of a single trapped $^{40}$Ca$^{+}$ ion. The ion is laser cooled and trapped in a linear Paul trap. The QNG states are prepared and analyzed are on the axial motional mode with a frequency of $\omega_z= 2 \pi\times 1.2$~MHz. The experimental sequence begins with the cooling of the ion to the motional ground state by Doppler cooling on the $4{\rm S}_{1/2}  \leftrightarrow 4{\rm P}_{1/2}$ transition followed by sideband cooling on the $|{\rm g}\rangle=4{\rm S}_{1/2}(m=-1/2) \leftrightarrow |{\rm e}\rangle=3{\rm D}_{5/2}(m=-5/2)$ transition. At the end of the cooling sequence the motional population of the vacuum state corresponds to $P_0=0.97\pm 0.02$ and the electronic state corresponds to $|{\rm g}\rangle$. In the next part of the sequence, the Fock states of motion are prepared by alternately applying blue and red sideband $\pi$-pulses~\cite{leibfried1996experimentalS, roos1999quantumS,mccormick2019quantumS}. The method employs the most fundamental and accessible interactions available within the spin-motion coupling on trapped ions corresponding to a Jaynes-Cummings and anti-Jaynes-Cummings interactions and can be realized with high fidelity allowing for the realization of unprecedentedly high motional Fock states~\cite{mccormick2019quantumS}. We note, that several alternative methods for the preparation of Fock states of motion of single ions have been proposed~\cite{cirac1993preparationS, eschner1995stochasticS, cirac1994nonclassicalS,blatt1995trappingS,myatt2000decoherenceS} and implemented~\cite{podhora2020unconditionalS,wolf2019motionalS,um2016phononS,kienzler2015quantumS}, which suggest further potential for enhancement of the generated state properties for high Fock states or implementations without the necessity of initialization of the oscillator in the motional ground state. Prior to the detection of the motional populations, a short 854~nm reshuffling pulse followed by the optical pumping are applied to transfer the residual population from $|e\rangle$ to the electronic ground state. To suppress the probability of the recoil in this process we apply an additional carrier $\pi$-pulse in the sequences for generation of the odd number states, where dominant electronic population after $n$-th population-raising pulse resides in the excited state $|e\rangle$.

\subsection{State detection and evaluation}
\label{stateDetection}

The detection of the phonon number distribution is performed by the measurement of the Rabi oscillation on the blue motional sideband of $|g\rangle \leftrightarrow |e\rangle$ transition. The phonon number distribution of the generated state is estimated by fitting the measured Rabi oscillations using the model
\begin{equation}
P_{\rm g}(t) = \frac{1}{2}(1-\sum_{n}^{n_{\rm max}} P(n) \cos(\Omega_{n} t) \exp^{-\gamma(n)t}),
\label{eq:RabiOscillationFit}
\end{equation}
where $\Omega_{n}= \sqrt{n+1} \Omega_{\rm c}  \eta $ is the Rabi frequency of the interaction on the state with $n$-phonons. The Rabi frequency on the corresponding carrier transition $\Omega_{\rm c}=2 \pi\times 69.7\pm 0.1$~kHz and Lamb-Dicke parameter $\eta = 0.0632\pm 0.0002$ were estimated from an independent set of measurements of Rabi oscillations on the carrier transition and on the first blue motional sideband for the ion prepared in the motional ground state. $\gamma_n$ corresponds to the decay parameter accounting mostly for the 729~nm laser intensity noise and beam pointing drifts. It scales with $n$ as~$\gamma_n = (n+1)^x \gamma_0$, where $\gamma_0=2\pi\times (0.054\pm 0.008)$~kHz was estimated from the observed decay of Rabi oscillations on the motional ground state. To limit any bias effect on the estimated statistics, we have devoted a significant attention to correct determination and setting of the factor $x$, which sets the scaling of the damping of Rabi oscillations with $n$~\cite{meekhof1996generationS}. We have investigated the dominant source of the Rabi flops decay by quantitative evaluation of observable Rabi oscillations for a range of prepared states approaching Fock states. This analysis showed, that up to a residual small decay caused by the population of other number states, the number of the observable Rabi oscillations and evolution of their contrast remained the same for all prepared states from $|0\rangle$ to $|10\rangle$. While according to the theory~\cite{schneider1998decoherenceS}, this confirms the dominant contribution of laser intensity fluctuations to the observed decay and suggests $x=1$, we have observed a highest-fidelity reproduction of the measured Rabi oscillations with a broadly employed value of $x\sim 0.7$, which can be likely attributed to the particular spectral characteristics of the noise~\cite{schneider1998decoherenceS,murao1998decoherence,bonifacio2000model,budini2002localization}. Despite the enormous progress in the manipulation of motional states of trapped ions since early experiments on a generation of nonclassical states~\cite{leibfried2003quantumS}, a robust and an efficient technique for unambiguous determination of $x$ remains an unresolved task.

\begin{figure*}[!t]
	\includegraphics[width=1.4\columnwidth]{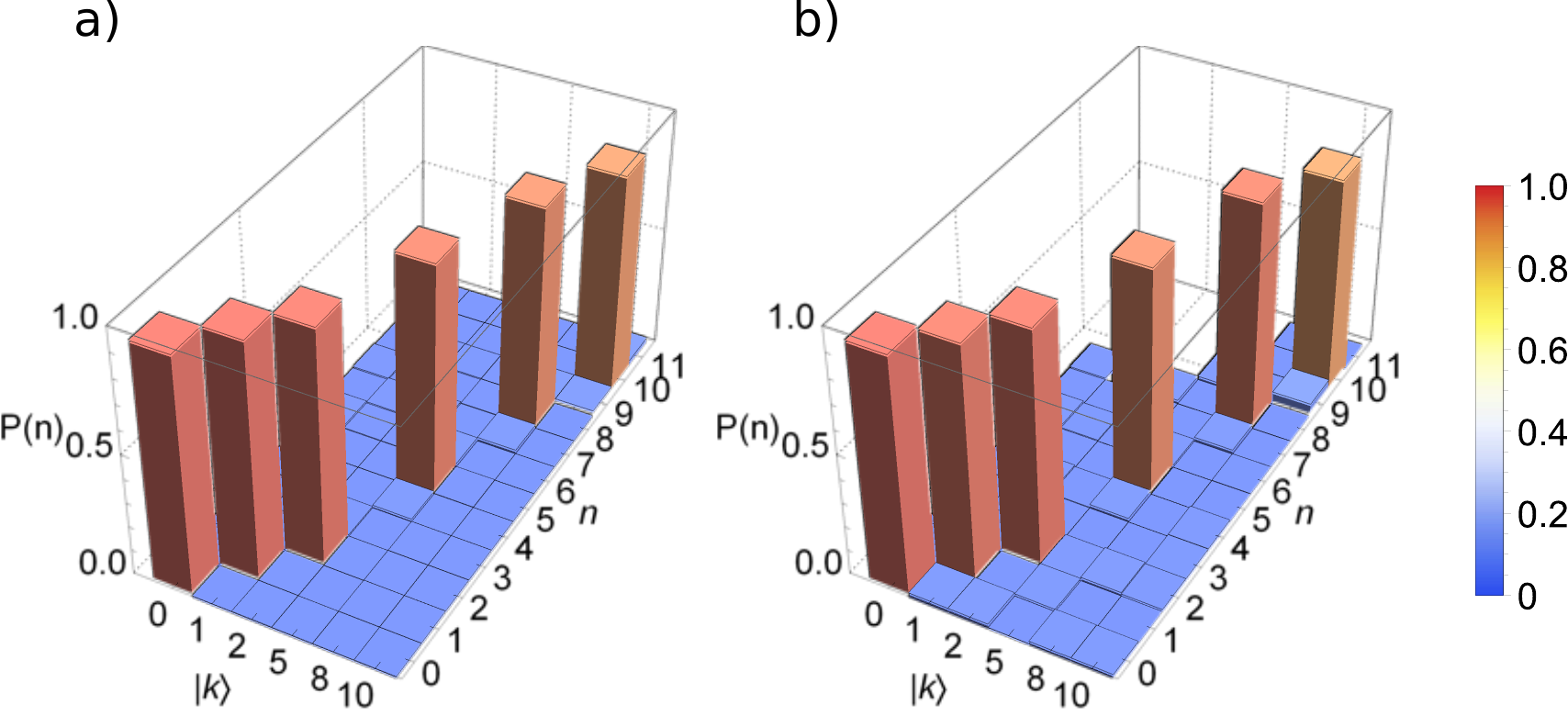}
	\caption{Phonon number distributions $p_n$ of the generated states approaching Fock states $|k\rangle$. Fig.~a) shows the simulated distributions taking into account the residual heating during state preparation, b) shows the measured phonon populations.}
	\label{fig:initStates3D}
\end{figure*}

The Rabi oscillations on the first blue motional sideband are measured with a time resolution such that even for the highest observed frequency corresponding to the state $|k_{\rm max}\rangle\sim |12\rangle$ we obtain more than 32 sample points per one Rabi oscillation period. Each sample point is the average of binary measurement results from 100~repetitions of the pulse sequence.

The measured Rabi oscillations are fitted without any prior knowledge about the observed probability distribution using the Eq.~\ref{eq:RabiOscillationFit} with a condition of non-negativity of probabilities~$P_n$. The resulting phonon-number probability distribution is normalized to unity and the maximal $n_{\rm max}$ in the fitting routine is chosen such that $n_{\rm max}=N+2$, where $N$ corresponds to the target Fock state $|N\rangle$. This suffices for effective coverage of the phonon-number distribution, as the probability of the population of higher number states in the employed state preparation sequence is negligible, which has been also confirmed by fits of Rabi oscillations with larger space of $n$.

\subsection{Generated states}
\label{SI:generatedStates}

Any quantum state preparation and detection procedure is in realistic experimental conditions influenced by noise. The dominant deteriorating effect in state preparation corresponds to the motional heating of the employed axial motional mode. The heating rate has been estimated to $\dot{\overline{n}}= 2.7\pm 0.2$~phonons/s for the employed motional frequency of $\omega_z = 2 \pi \times 1.2$~MHz. The residual small deviations in the presented generations of GQNG states correspond to the finite efficiency of population transfer in each of the Rabi $\pi$-pulses, which has been estimated to be very close to unity and can be mostly attributed to the residual laser intensity fluctuations within the interaction of 729~nm laser beam with the ion and beam pointing drifts.

\begin{figure*}[!t]
\includegraphics[width=1.4\columnwidth]{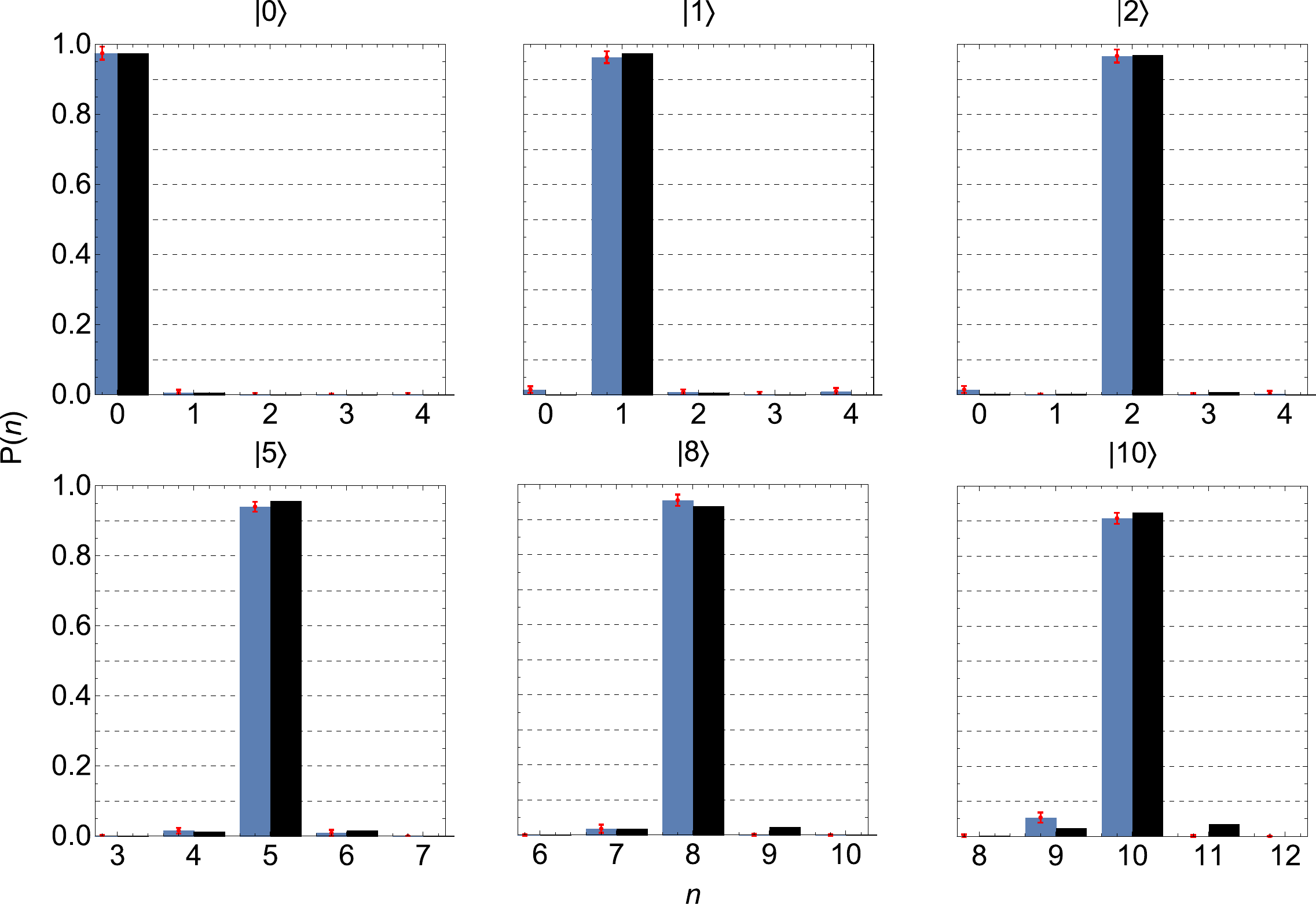}
\caption{Parts of the phonon-number distributions $P(n)$ of the generated states approaching Fock states $|k\rangle$ relevant for the estimation of the GQNG properties are depicted with blue bars with black bars corresponding to the simulation considering the imperfect solely heating of ion in the state generation process. The error bars correspond to a single standard deviation.}
\label{fig:initStates}
\end{figure*}

Fig.~\ref{fig:initStates3D} shows the measured phonon-number distributions for the states approaching Fock state $|k\rangle$ and corresponding simulation of the state resulting from the series of the $\pi$-pulses limited solely by the motional heating. The simulations quantitatively reproduce the measured statistics, with the average fidelity between statistical distributions of 0.96. The observed states populations confirm important aspects of the generation of GQNG states in a trapped-ion mechanical oscillator. They show that even in the case of extremely well controlled and isolated system represented by a single trapped ion, mechanical heating corresponds to a critical process determining the quality of the output state. In addition, for the employed procedure starting from an initial motional ground state, the observed populations resulting from the heating process and additional infidelities are concentrated in states $|n\rangle$, with $n\leq k+1$, where $|k\rangle$ is the target state. The~Fig.~\ref{fig:initStates} shows the sections of the presented distributions with populations concentrated around the target state relevant for the evaluated GQNG criteria. The error bars are estimated by a Monte-Carlo simulation from the projection noise in each data point of the measured blue-sideband Rabi oscillations corresponding to 100~projection measurements. The displayed uncertainties thus correspond to a lower bound on the uncertainty of $P_n$ and do not take into an account long-term drifts of experimental parameters. The analogous procedure for the estimation of the experimental uncertainties has been employed for all measured phonon-number distributions and evaluated parameters presented in this work. The only exception corresponds to the estimation of the carrier Rabi frequency $\Omega_{\rm c}$ and decay parameter $\gamma_0$. These were measured and estimated in independent sets of five measurements of Rabi oscillations on of the carrier transition for $\Omega_{\rm c}$ and on the first blue motional sideband for $\gamma_0$, respectively, and then evaluated statistically.

\section{Emulation of state thermalization}
\label{SI:thermalization}

\subsection{Theoretical model}

Here, we make a simple quantitative model accounting for the effect of the motional heating induced by the random photon recoils from the scattering of the 397~nm Doppler cooling laser beam. Close to the motional ground state, it is sufficient to describe the process as many phase-randomized displacements parameterized by a single parameter $\bar{n}$ accounting for the strength or length of the applied randomized displacements. However, when approaching the Doppler cooling limit, which has been estimated in the presented experiment to about $6.7\pm 0.7$ phonons, the dynamics corresponding to the interaction with a finite temperature reservoir has to be included which corresponds to a two-parameter characterization of the heating process. We note, that the simplified single-parameter thermalization dynamics can be used for universal estimation of the QNG depth, because it effectively maximizes the entropy of the output state for the given exchange of energy.

\emph{Ideal thermalization} -- The QNG depth is defined and characterized for each prepared Fock state to characterize the quality of generated nonclassical state and to test its robustness against thermalization mechanism caused by interaction with a thermal environment corresponding to a naturally dominant and well characterized state-deteriorating process for mechanical oscillators~\cite{brownnutt2015ionS,aspelmeyer2014cavityS}. The interaction is emulated by a short excitation on a dipole transition, which precedes the state analysis using a laser set to the Doppler cooling conditions. The process effectively corresponds to the implementation of displacements in phase space with an amplitude $\alpha$ that change its value randomly according to the Gaussian distribution. The interaction with the thermal environment effectively deteriorates the initial state $\rho_i$ according to the map
\begin{equation}
M_{\bar{n}}(\rho_i) = \int d^2 \alpha e^{-\frac{|\alpha|^2}{\bar{n}}} D(\alpha)\rho_{i}D^\dagger(\alpha).
\label{eqS:heatingFull}
\end{equation}
For very small $\bar{n}$, this can be further simplified to
\begin{equation}
M_{\bar{n}}(\rho_i)\approx \rho_i+\bar{n}^2[a \rho a^{\dagger}+a^{\dagger}\rho a-(a^{\dagger} a+1/2)\rho_i-\rho_i(a^{\dagger} a+1/2),
\label{eqS:heating}
\end{equation}
where terms proportional to $\bar{n}^4$ are neglected. We note that all terms in the first order of annihilation and creation operators disappear due to averaging over the phase of a random displacement. The map~$M_{\bar{n}}$ in this approximation represents a deterioration of the motion caused by scattering of a single photon on the ion with $\bar{n}$ being interpreted as the Lamb-Dicke parameter. The map in Eq.~(\ref{eqS:heatingFull}) can be employed for evaluation of depth of the GQNG, which is determined by the maximal parameter $\bar{n}$ preserving its observability.

\textit{Thermalization close to the Doppler cooling limit} -- The acting of the map $M_{\bar{n}}$ represents a good approximation of the description of the motional evolution of ion close to the motional ground state. However, close to the laser cooling limit corresponding to the applied red-detuned 397~nm thermalization beam, the motional dynamics is appropriately described by rate equations assuming damping of the state $|n\rangle$ to $|n-1\rangle$ with the rate $A$ or excitation of $|n\rangle$ to $|n+1\rangle$ with the rate $B$~\cite{eschner2003laserS}. The evolution of the motional states obeys the differential equations
\begin{equation}
\frac{d}{d t} P_{n} = A(n+1)P_{n+1} + B n P_{n-1} - [A n + B(n+1)]P_{n}
\label{eq:DopplerHeating}
\end{equation}
for every $n\geq0$ with $t$ corresponding to the interaction time. The Bose-Einstein statistics of a motional state with $P_n=1/(1+\bar{n}_{\rm BE})\left(\bar{n}_{\rm BE}/(1+\bar{n}_{\rm BE})\right)^n$, where $\bar{n}_{BE}=B/(A-B)$ represents the steady state solution of equation~(\ref{eq:DopplerHeating}) for $A>B$. The opposite case with $A<B$ does not enable achieving the steady state solution since the heating process causes divergence of the mean phonon number for a large interaction time. Moreover, the equations~(\ref{eq:DopplerHeating}) preserve the Bose-Einstein statistics, which results in their simplification to a single equation for the mean number of phonons $\bar{n}_{\rm BE}$
\begin{equation}
\frac{d \bar{n}_{\rm BE}}{d t} =A \bar{n}_{\rm BE} - B(1+\bar{n}_{\rm BE})
\end{equation}
Beyond this simple example, the Eqs.~\ref{eq:DopplerHeating} can be solved only numerically. Advantageously, formulas for states after a short interaction time can be derived approximately with certain accuracy. In this limit, the state resulting from the thermalization of the initial number state $|m\rangle$ is given by the incoherent sum of probabilities $P_{n}$ corresponding to the Taylor series
\begin{equation}
P_n=\sum_{k=\vert m-n\vert}^{\infty}\mu_{n,k} t^k.
\label{pNDoppler}
\end{equation}
where $t$ stands for the interaction time. It allows us to express the derivative $\frac{d}{d t} P_n$ and compare it with the right-hand side of Eq.~\ref{eq:DopplerHeating}. Since it gives rise to identities holding for any $t$, the coefficients $\mu_{m,k}$ with $k$ up to some chosen index $N$ can be determined by comparing members with the same power of $t$. The higher index $N$ the better accuracy of the solution is achieved.

\begin{figure*}[!t]
\includegraphics[width=2\columnwidth]{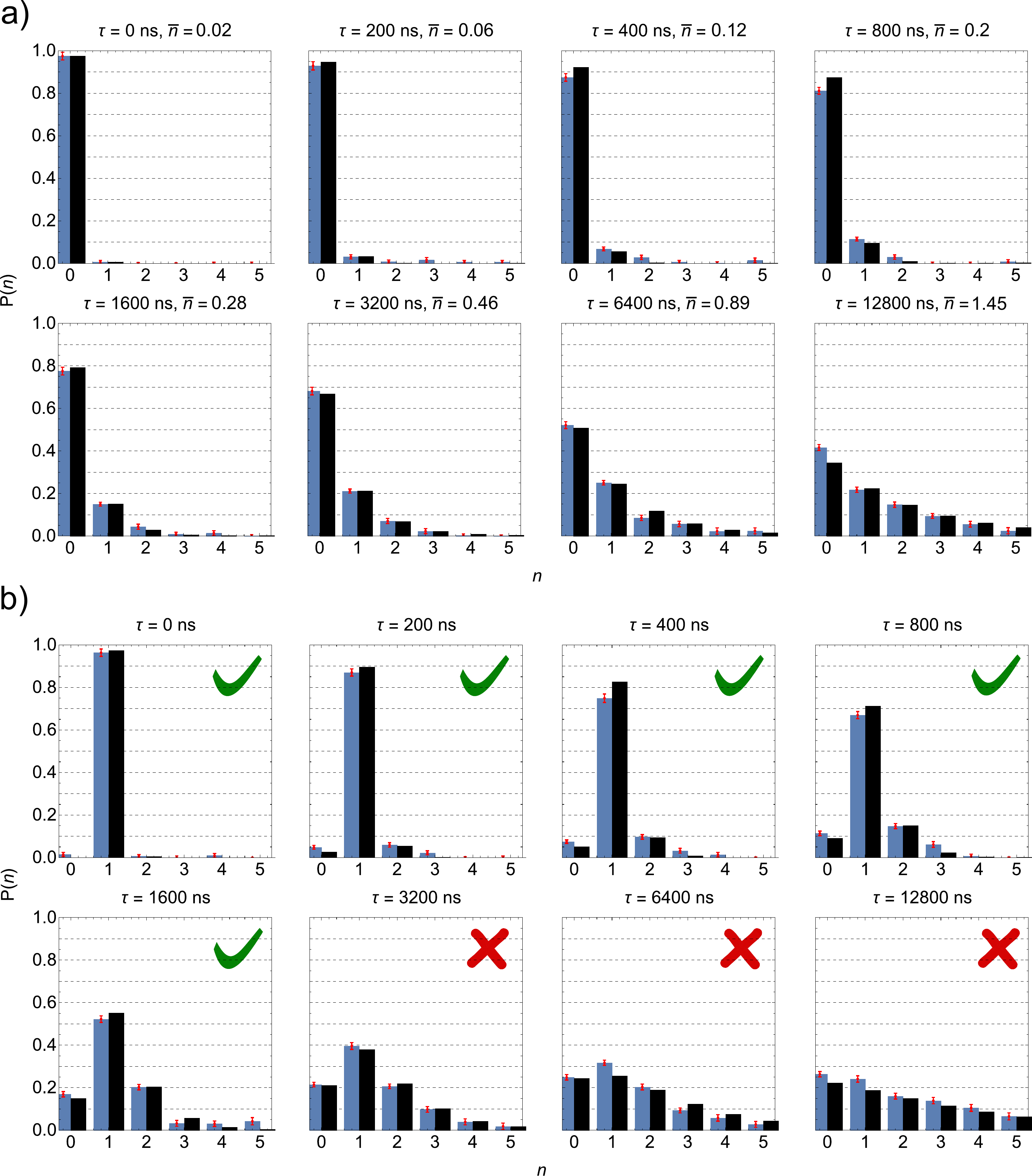}
\caption{Measured (blue bars) and simulated (black bars) phonon number distributions undergoing thermalization for various durations of the 397~nm thermalization laser pulse $\tau$. The statistics in a) and b) show the evolution for the motional ground state $|0\rangle$ and state approaching Fock state $|1\rangle$, respectively. The error bars corresponding to a single standard deviation were estimated by a Monte Carlo simulation of errors resulting from projection noise in the measured Rabi oscillations.}
\label{fig:Fock0_therm}
\end{figure*}

In contrast to the previous model where the process aggravating the phonon-distribution is characterized by a single parameter $\bar{n}$, the Doppler heating is determined by the time of the interaction and by the ratio of coefficients $A/B$. Thus, the analysis of the genuine $n$-photon quantum non-Gaussian depth becomes more complex when the thermalization together with the full Doppler cooling dynamics is considered.

\subsection{Experimental implementation}

The depth of the realized GQNG and its robustness to the realistic noise have been experimentally accessed by the estimation of its observability after the photon-recoil induced thermalization pulse. Thermalization with controllable speed and duration is implemented by photon scattering on the 4S$_{1/2} \leftrightarrow {\rm 4P}_{1/2}$ transition using a short 397~nm laser pulse together with the 866~nm laser which reshuffles the population of the metastable ${\rm 3D}_{3/2}$ state back to the cooling transition. The 397~nm laser detuning and power have been set so that they correspond to our standard Doppler-cooling settings. The thermalization pulse duration has been ranging from 100~ns up to approximately 12.8~$\mu$s, which corresponds to the addition of $n_{\rm th}=0.02\pm 0.01$ and $1.45\pm 0.04$ of mean thermal quanta, respectively, when applied to the motional ground state. The probability of the photon recoil and the corresponding speed of thermalization effect increase for higher motional states.

Prolonging of the thermalization pulse leads to the gradual diffusion of the population $p_k$ into neighboring oscillator states. In the limit of infinite 397~pulse length, the statistics approaches the ideal thermal state. For the number states $|k\rangle$ with $k\leq 10$ presented in this manuscript, this limit is reached by the thermalization pulse lengths on the timescales of several milliseconds.

Fig.~\ref{fig:Fock0_therm}-a) shows the measurement of the evolution of the phonon-number distribution for several lengths of the thermalization pulse $\tau$ applied to the motional ground state $|0\rangle$. The evolution of the state populations reproduces the expected distributions given by the model considering the phase-randomized photon recoils described in the previous subsection~\ref{SI:thermalization}-A, with a rate of addition of a mean thermal energy following the corresponding simulated scaling from the Eq.~\ref{eqS:heating}. The fundamental role of the initial vacuum state together with the predictability and single-parameter dependence of the presented generally implementable heating mechanism corresponding to the recoil heating provide a reproducible and broadly applicable way of comparison of the robustness of the generated GQNG states between different experimental platforms.

Fig.~\ref{fig:Fock0_therm}-b) shows the results of the thermalization process applied to the state approaching the Fock state $|1\rangle$. It illustrates the thermal diffusion of the initial statistical distribution with progressing transfer of the population into the nearest number states. The presented data confirm that, while in the short thermalization times the populations of $|0\rangle$ and $|2\rangle$ are nearly symmetric with a slight preference of the population of $|2\rangle$ due to the relative amplitudes of annihilation and creation operations in the Eq.~\ref{eqS:heating}, the diffusion of $|2\rangle$ to higher number states for longer thermalization times reverses this slight asymmetry. Importantly, the GQNG properties can be unambiguously identified for the presented statistics up to the thermalization pulse lengths $\tau=1.28\,\mu$s and the measured phonon number distributions agree with the simulated statistics to the level of the estimated experimental uncertainties shown as a single standard deviation.

\section{Wigner function}
\label{SI:Wigner}

\begin{figure}[!t]
\includegraphics[width=\columnwidth]{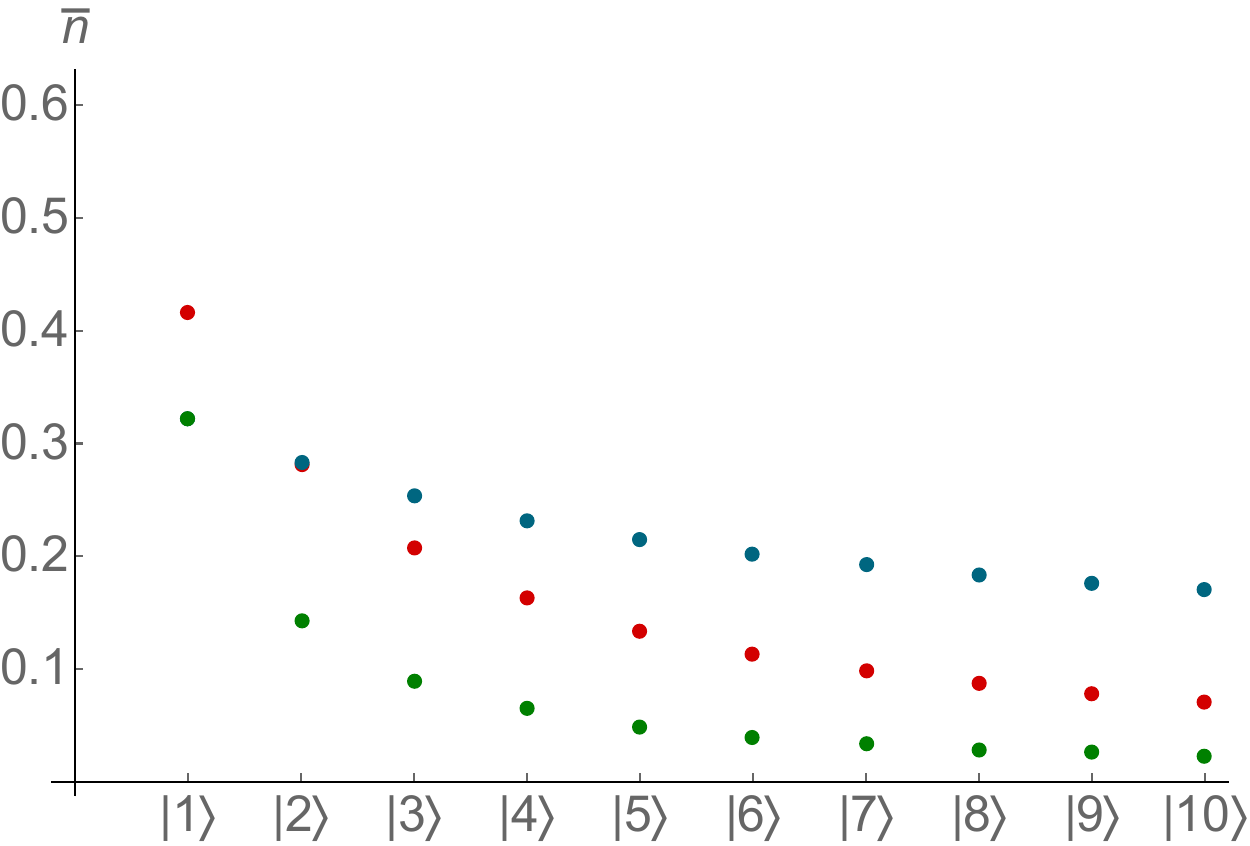}
\caption{The thermal depth of the genuine $n$-phonon quantum non-Gaussianity (green points), the quantum non-Gaussianity (blue points) and negativity in the Wigner function (red points) that are exhibited by the ideal Fock states. The thermalization deteriorates the Fock states according to the map (\ref{eqS:heatingFull}). The vertical axis quantifies the maximal mean number of thermal phonons that preserves the presented quantum aspects. Note, the quantum non-Gaussianity and genuine one-phonon quantum non-Gaussianity are identical properties, and therefore their depth is the same for $|1\rangle$.}
\label{fig:Wig}
\end{figure}

The genuine $n$-phonon quantum non-Gaussianity of pure states manifests itself by the proper number of negative annuli in the Wigner function. The topology of negative regions in the Wigner function exposes the genuine $n$-order quantum non-Gaussianity because each Fock state exhibits a specific number of annuli, which is not changed by the squeezing or displacement. Note, that the reliability of such criterion holds straightforwardly only for pure states since stochastic processes affecting the states $D(\alpha) S(r) \sum_{m=0}^{n-1} c_{m} |n\rangle$ can partially increase the number of negativities.
To show the negativity can be misleading for that recognition, we compare the thermal depth of the genuine $n$-phonon quantum non-Gaussianity manifested by the probability $P_n=\langle n|\rho|n\rangle$ with the thermal depth that the negativity in the Wigner function determines. The comparison is dependent on a specific model of thermalization and state preparation. Therefore, we consider ideal Fock states, which are deteriorated according to the map (\ref{eqS:heatingFull}). The Fig.~\ref{fig:Wig} depicts the depth of the genuine $n$-phonon QNG and the negativity in the Wigner function. Note, the thermalization reduces significantly all the negativities before they disappear completely. In reality, the statistical errors will however matter and negative values can also appear as an artefact of numerical processing. Thus, we required in the analysis that all the negative peaks in the Wigner function achieves at least $5\%$ of their value before the thermalization. Nevertheless, the negativity survives very strong noise contributions compared to the genuine $n$-phonon QNG, which hints that the negativity in the Wigner function is limited for certification of the genuine $n$-phonon quantum non-Gaussianity for strongly thermalized states.

\section{Force estimation capability}
\label{SI:sensing}

Fock states can be advantageously employed for the sensing of a force exerting weak displacement in the motion of the initial state $\rho$, which yields the displaced state
\begin{equation}
    \rho_{|\alpha|^2}=\frac{1}{2\pi} \int \mathrm{d}\phi D(|\alpha|\exp(i \phi))\rho D^{\dagger}(|\alpha|\exp(-i \phi)),
\end{equation}
where $D(|\alpha|\exp(i \phi))=\exp\left(|\alpha|^2\exp(i \phi) a^{\dagger}-|\alpha|\exp(-i \phi) a\right)$ is the displacement operator and $|\alpha|^2$ quantifies the mean number of phonons that the displacement induces. Information acquired in the sensing of $|\alpha|^2$ is quantified by the Fisher information
\begin{equation}
F=\sum_{m=0}^{\infty}\frac{1}{P_m(|\alpha|^2)}\left[\frac{\mathrm{d}}{\mathrm{d}{|\alpha|^2}} P_m(|\alpha|^2)\right]^2,
\label{eq:fisher}
\end{equation}
where $P_m(|\alpha|^2)=\langle m|\rho_{|\alpha|^2}|m\rangle$. The operator $D(|\alpha|)$ acting on the Fock states induces the phonon distribution given by
\begin{equation}
    |\langle n|D(|\alpha|)| m\rangle|^2=\frac{n!}{m!}e^{-|\alpha|^2}\sum_{k=0}^{m}\frac{(-|\alpha|^2)^k}{(k+n-m)!}{m \choose k}
    \label{dF1}
\end{equation}
or
\begin{equation}
    |\langle n|D(|\alpha|)| m\rangle|^2=\frac{m!}{n!}e^{-|\alpha|^2}\sum_{k=0}^{n}\frac{(-|\alpha|^2)^k}{(k+m-n)!}{n \choose k}
    \label{dF2}
\end{equation}
for $m\leq n$ or $n\leq m$, respectively. Equipped with identities (\ref{dF1}) and (\ref{dF2}), the Fisher information of the state $\rho_{|\alpha|^2}$ follows
\begin{equation}
    F=\sum_{n=0}^{\infty}\sum_{m=0}^{\infty} \frac{1}{\rho_m \delta_{n,m}(|\alpha|^2)}\left[\rho_m \frac{\rm d}{{\rm d}_{|\alpha|^2}} \delta_{n,m}(|\alpha|^2)\right]^2,
\end{equation}
where $\delta_{n,m}=|\langle n|D(|\alpha|)| m\rangle|^2$ and $\rho_m=\langle m |\rho| m \rangle$ with $\rho$ corresponding to initial density matrix of the motional state before performing the displacement.

\begin{figure}[!t]
\includegraphics[width=\columnwidth]{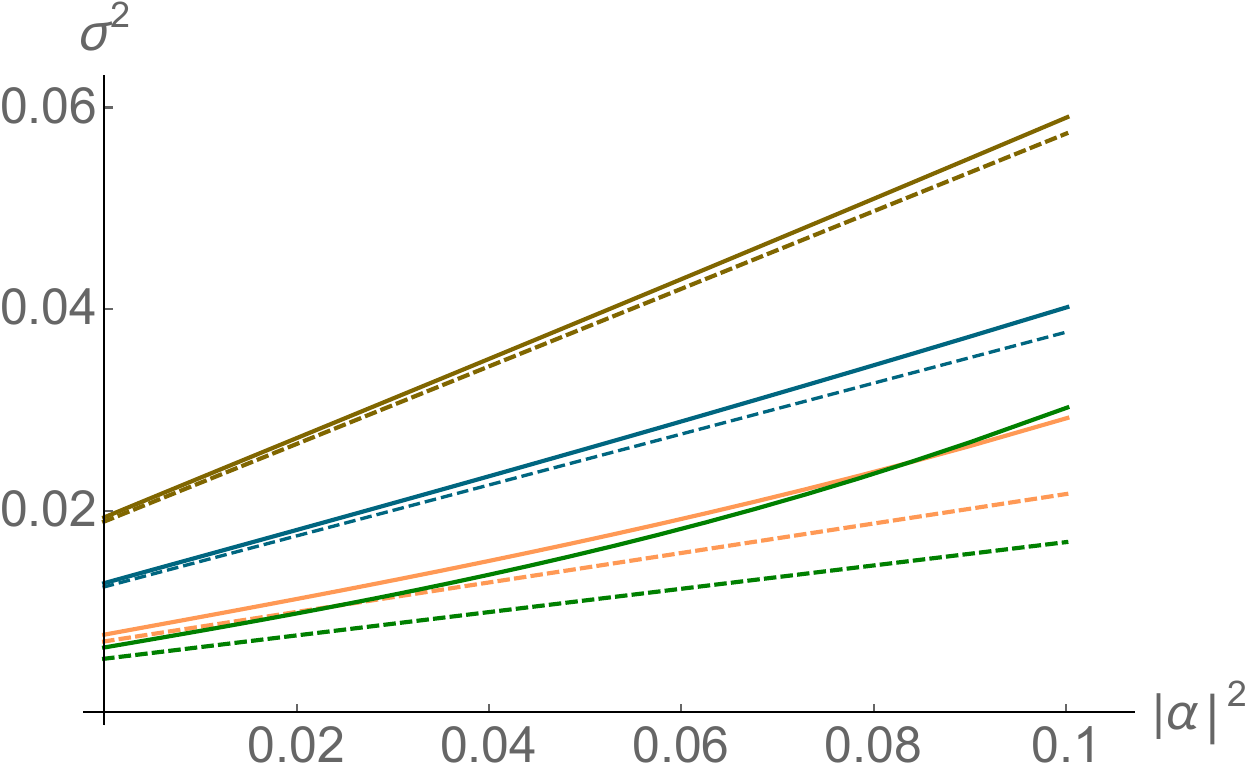}
\caption{Demonstration of how the approximation (\ref{sigma}) converges to the estimation error given from the true value of the Fisher information. The model states are Fock states that are affected by the thermalization process according to the map (\ref{eqS:heatingFull}) with $\bar{n}=0.05$ before they are exploited in the sensing the displacement $|\alpha|^2$. The depicted estimation error $\sigma^2$ induces the noisy Fock state $|1\rangle$ (yellow), $|2\rangle$ (blue), $|5\rangle$ (orange) and $|8\rangle$ (green). Whereas the solid lines stand for the true $\sigma^2$, the dashed line represents the approximation (\ref{sigma}).}
\label{fig:check}
\end{figure}

For ideal detector distinguishing the individual phonons, the estimation error of $|\alpha|^2$ is $\sigma^{2}= \frac{1}{F N}$ with $N$ being the number of samples.
Performing the sensing with the ideal Fock state $|n\rangle$ allows us to express the estimation error according to
\begin{equation}
1/\sigma^2=\sum_{m=0}^{\infty}\frac{\left[\frac{\mathrm{d}}{\mathrm{d}|\alpha|^2}|\langle n|D(|\alpha|)| m\rangle|^2\right]^2}{|\langle n|D(|\alpha|)| m\rangle|^2}.
\label{sigmaFock}
\end{equation}
We can simplify the formula (\ref{sigmaFock}) for sensing of the small $|\alpha|^2$ by expanding it into Taylor series. We confirmed the series obeys
\begin{equation}
    \sigma^{2}=\frac{|\alpha|^2}{(2n +1) N}+o^{(4)}(|\alpha|^2),
\end{equation}
where $o^{(4)}$ refers to some function such that $\lim_{|\alpha|^2\rightarrow 0}o^{(4)}(|\alpha|^2)/|\alpha|^6=0$.
Thus, the Fock state $|n\rangle$ reduces the estimation error faster than all the lower Fock states. Note, the realistic Fock states always deviate from the ideal Fock states, which increase the estimation error $\sigma^{2}$ in the real situations. To demonstrate that, let us allow for the state $\rho_{\bar{n}}$ representing a density matrix approaching the Fock state $|n\rangle \langle n|$ that is affected weakly by the thermal environment according to the map (S7)
\begin{equation}
    \begin{aligned}
        \rho_{\bar{n}} &\approx P_n |n\rangle \langle n|+n \bar{n}|n-1\rangle \langle n-1|\\
        &+(n+1) \bar{n}|n+1\rangle \langle n+1|,
    \end{aligned}
    \label{approxSt}
\end{equation}
where $\bar{n} \ll 1$ is a mean number of added phonons and $\rho_n=1-\bar{n}(2n+1)$ is the probability of $n$ phonons in motional state. Let us consider a measurement distinguishing the presence of $n-1$ phonons, $n$ phonons, $n+1$ phonons. Acting the displacement operator on the state (\ref{approxSt}) produces the probabilities
\begin{equation}
    \begin{aligned}
    P_{n-1}\left(|\alpha|^2\right)&\approx \rho_n n |\alpha|^2+n \bar{n} \left[1-\left(2n-1\right)\right]|\alpha|^2\nonumber \\
    P_{n}\left(|\alpha|^2\right)&\approx \rho_n \left[1-\left(2n+1\right)|\alpha|^2\right]\\
    &+\bar{n} \left[\left(n+1\right)^2+n^2\right]|\alpha|^2\nonumber \\
    P_{n+1}\left(|\alpha|^2\right)&\approx \rho_n (n+1) |\alpha|^2\\
    &+(n+1) \bar{n} \left[1-\left(2n+3\right)\right]|\alpha|^2,
    \end{aligned}
\end{equation}
where the Taylor series of (\ref{dF1}) or (\ref{dF2}) with respect to $|\alpha|^2$ is exploited. The Fisher information for the measurement used for sensing of $|\alpha|^2$ achieves
\begin{equation}
\begin{aligned}
    F&=\sum_{k=n-1}^{n+1}\frac{1}{P_k\left(|\alpha|^2\right)}\left[\frac{d P_k\left(|\alpha|^2\right)}{d |\alpha|^2}\right]^2\\
    &+\frac{1}{1-\sum_{k=n-1}^{n+1}P_k\left(|\alpha|^2\right)}\left[\frac{d }{d |\alpha|^2}\sum_{k=n-1}^{n+1} P_k\left(|\alpha|^2\right)\right]^2.
\end{aligned}
\label{Fapp}
\end{equation}
Further, we approximate the estimation error $\sigma^2$ saturating the Fisher information (\ref{Fapp}) by expanding it in Taylor series for very small $|\alpha|^2$. To achieve a simple formula with clear interpretation, we introduce the probability $P_e=\langle n-1|\rho_{\bar{n}}|n-1\rangle+\langle n+1|\rho_{\bar{n}}|n+1\rangle$ and expand further $\sigma^2$ in Taylor series with respect to $P_e$, which works out to be
\begin{equation}
    \sigma^{2}\approx \frac{|\alpha|^2}{(2n +1)P_n N}+\frac{P_e}{(2n +1)^2 N P_n^2}.
    \label{sigma}
\end{equation}
Thus, estimation error (\ref{sigma}) is given by the member that is proportional to $|\alpha|^2$, which remain relevant even for the ideal Fock states, and the member independent of $|\alpha|^2$ and proportional to $P_e$, which is responsible for growth of (\ref{sigma}) for small $|\alpha|^2$ due to the noise in the intial state. In practice, therefore these absolute offset by second term is always present, however, it can be reduced if the Fock state quality remains same with increasing $n$. Fig.~\ref{fig:check} compares the approximation (\ref{sigma}) with the accurate solution for the estimation error $\sigma^2$. It demonstrates the approximation (\ref{sigma}) is limited to small $|\alpha|^2$ and to low $n$ of the noisy Fock states in (\ref{approxSt}) as well.

\begin{thebibliography}{61}%
\makeatletter
\providecommand \@ifxundefined [1]{%
 \@ifx{#1\undefined}
}%
\providecommand \@ifnum [1]{%
 \ifnum #1\expandafter \@firstoftwo
 \else \expandafter \@secondoftwo
 \fi
}%
\providecommand \@ifx [1]{%
 \ifx #1\expandafter \@firstoftwo
 \else \expandafter \@secondoftwo
 \fi
}%
\providecommand \natexlab [1]{#1}%
\providecommand \enquote  [1]{``#1''}%
\providecommand \bibnamefont  [1]{#1}%
\providecommand \bibfnamefont [1]{#1}%
\providecommand \citenamefont [1]{#1}%
\providecommand \href@noop [0]{\@secondoftwo}%
\providecommand \href [0]{\begingroup \@sanitize@url \@href}%
\providecommand \@href[1]{\@@startlink{#1}\@@href}%
\providecommand \@@href[1]{\endgroup#1\@@endlink}%
\providecommand \@sanitize@url [0]{\catcode `\\12\catcode `\$12\catcode
  `\&12\catcode `\#12\catcode `\^12\catcode `\_12\catcode `\%12\relax}%
\providecommand \@@startlink[1]{}%
\providecommand \@@endlink[0]{}%
\providecommand \url  [0]{\begingroup\@sanitize@url \@url }%
\providecommand \@url [1]{\endgroup\@href {#1}{\urlprefix }}%
\providecommand \urlprefix  [0]{URL }%
\providecommand \Eprint [0]{\href }%
\providecommand \doibase [0]{https://doi.org/}%
\providecommand \selectlanguage [0]{\@gobble}%
\providecommand \bibinfo  [0]{\@secondoftwo}%
\providecommand \bibfield  [0]{\@secondoftwo}%
\providecommand \translation [1]{[#1]}%
\providecommand \BibitemOpen [0]{}%
\providecommand \bibitemStop [0]{}%
\providecommand \bibitemNoStop [0]{.\EOS\space}%
\providecommand \EOS [0]{\spacefactor3000\relax}%
\providecommand \BibitemShut  [1]{\csname bibitem#1\endcsname}%
\let\auto@bib@innerbib\@empty
\bibitem [{\citenamefont {Leibfried}\ \emph {et~al.}(1996)\citenamefont
  {Leibfried}, \citenamefont {Meekhof}, \citenamefont {King}, \citenamefont
  {Monroe}, \citenamefont {Itano},\ and\ \citenamefont
  {Wineland}}]{leibfried1996experimental}%
  \BibitemOpen
  \bibfield  {author} {\bibinfo {author} {\bibfnamefont {D.}~\bibnamefont
  {Leibfried}}, \bibinfo {author} {\bibfnamefont {D.}~\bibnamefont {Meekhof}},
  \bibinfo {author} {\bibfnamefont {B.}~\bibnamefont {King}}, \bibinfo {author}
  {\bibfnamefont {C.}~\bibnamefont {Monroe}}, \bibinfo {author} {\bibfnamefont
  {W.~M.}\ \bibnamefont {Itano}},\ and\ \bibinfo {author} {\bibfnamefont
  {D.~J.}\ \bibnamefont {Wineland}},\ }\href
  {https://journals.aps.org/prl/abstract/10.1103/PhysRevLett.77.4281}
  {\bibfield  {journal} {\bibinfo  {journal} {Phys. Rev. Lett.}\ }\textbf
  {\bibinfo {volume} {77}},\ \bibinfo {pages} {4281} (\bibinfo {year}
  {1996})}\BibitemShut {NoStop}%
\bibitem [{\citenamefont {Meekhof}\ \emph {et~al.}(1996)\citenamefont
  {Meekhof}, \citenamefont {Monroe}, \citenamefont {King}, \citenamefont
  {Itano},\ and\ \citenamefont {Wineland}}]{meekhof1996generation}%
  \BibitemOpen
  \bibfield  {author} {\bibinfo {author} {\bibfnamefont {D.}~\bibnamefont
  {Meekhof}}, \bibinfo {author} {\bibfnamefont {C.}~\bibnamefont {Monroe}},
  \bibinfo {author} {\bibfnamefont {B.}~\bibnamefont {King}}, \bibinfo {author}
  {\bibfnamefont {W.~M.}\ \bibnamefont {Itano}},\ and\ \bibinfo {author}
  {\bibfnamefont {D.~J.}\ \bibnamefont {Wineland}},\ }\href
  {https://journals.aps.org/prl/abstract/10.1103/PhysRevLett.76.1796}
  {\bibfield  {journal} {\bibinfo  {journal} {Phys. Rev. Lett.}\ }\textbf
  {\bibinfo {volume} {76}},\ \bibinfo {pages} {1796} (\bibinfo {year}
  {1996})}\BibitemShut {NoStop}%
\bibitem [{\citenamefont {Roos}\ \emph {et~al.}(1999)\citenamefont {Roos},
  \citenamefont {Zeiger}, \citenamefont {Rohde}, \citenamefont {N{\"a}gerl},
  \citenamefont {Eschner}, \citenamefont {Leibfried}, \citenamefont
  {Schmidt-Kaler},\ and\ \citenamefont {Blatt}}]{roos1999quantum}%
  \BibitemOpen
  \bibfield  {author} {\bibinfo {author} {\bibfnamefont {C.}~\bibnamefont
  {Roos}}, \bibinfo {author} {\bibfnamefont {T.}~\bibnamefont {Zeiger}},
  \bibinfo {author} {\bibfnamefont {H.}~\bibnamefont {Rohde}}, \bibinfo
  {author} {\bibfnamefont {H.}~\bibnamefont {N{\"a}gerl}}, \bibinfo {author}
  {\bibfnamefont {J.}~\bibnamefont {Eschner}}, \bibinfo {author} {\bibfnamefont
  {D.}~\bibnamefont {Leibfried}}, \bibinfo {author} {\bibfnamefont
  {F.}~\bibnamefont {Schmidt-Kaler}},\ and\ \bibinfo {author} {\bibfnamefont
  {R.}~\bibnamefont {Blatt}},\ }\href
  {https://journals.aps.org/prl/abstract/10.1103/PhysRevLett.83.4713}
  {\bibfield  {journal} {\bibinfo  {journal} {Phys. Rev. Lett.}\ }\textbf
  {\bibinfo {volume} {83}},\ \bibinfo {pages} {4713} (\bibinfo {year}
  {1999})}\BibitemShut {NoStop}%
\bibitem [{\citenamefont {Toyoda}\ \emph {et~al.}(2015)\citenamefont {Toyoda},
  \citenamefont {Hiji}, \citenamefont {Noguchi},\ and\ \citenamefont
  {Urabe}}]{toyoda2015hong}%
  \BibitemOpen
  \bibfield  {author} {\bibinfo {author} {\bibfnamefont {K.}~\bibnamefont
  {Toyoda}}, \bibinfo {author} {\bibfnamefont {R.}~\bibnamefont {Hiji}},
  \bibinfo {author} {\bibfnamefont {A.}~\bibnamefont {Noguchi}},\ and\ \bibinfo
  {author} {\bibfnamefont {S.}~\bibnamefont {Urabe}},\ }\href
  {https://www.nature.com/articles/nature15735} {\bibfield  {journal} {\bibinfo
   {journal} {Nature}\ }\textbf {\bibinfo {volume} {527}},\ \bibinfo {pages}
  {74} (\bibinfo {year} {2015})}\BibitemShut {NoStop}%
\bibitem [{\citenamefont {Kienzler}\ \emph {et~al.}(2017)\citenamefont
  {Kienzler}, \citenamefont {Lo}, \citenamefont {Negnevitsky}, \citenamefont
  {Fl{\"u}hmann}, \citenamefont {Marinelli},\ and\ \citenamefont
  {Home}}]{kienzler2017quantum}%
  \BibitemOpen
  \bibfield  {author} {\bibinfo {author} {\bibfnamefont {D.}~\bibnamefont
  {Kienzler}}, \bibinfo {author} {\bibfnamefont {H.-Y.}\ \bibnamefont {Lo}},
  \bibinfo {author} {\bibfnamefont {V.}~\bibnamefont {Negnevitsky}}, \bibinfo
  {author} {\bibfnamefont {C.}~\bibnamefont {Fl{\"u}hmann}}, \bibinfo {author}
  {\bibfnamefont {M.}~\bibnamefont {Marinelli}},\ and\ \bibinfo {author}
  {\bibfnamefont {J.~P.}\ \bibnamefont {Home}},\ }\href
  {https://journals.aps.org/prl/abstract/10.1103/PhysRevLett.119.033602}
  {\bibfield  {journal} {\bibinfo  {journal} {Phys. Rev. Lett.}\ }\textbf
  {\bibinfo {volume} {119}},\ \bibinfo {pages} {033602} (\bibinfo {year}
  {2017})}\BibitemShut {NoStop}%
\bibitem [{\citenamefont {Ding}\ \emph {et~al.}(2017)\citenamefont {Ding},
  \citenamefont {Maslennikov}, \citenamefont {Habl{\"u}tzel}, \citenamefont
  {Loh},\ and\ \citenamefont {Matsukevich}}]{ding2017quantum}%
  \BibitemOpen
  \bibfield  {author} {\bibinfo {author} {\bibfnamefont {S.}~\bibnamefont
  {Ding}}, \bibinfo {author} {\bibfnamefont {G.}~\bibnamefont {Maslennikov}},
  \bibinfo {author} {\bibfnamefont {R.}~\bibnamefont {Habl{\"u}tzel}}, \bibinfo
  {author} {\bibfnamefont {H.}~\bibnamefont {Loh}},\ and\ \bibinfo {author}
  {\bibfnamefont {D.}~\bibnamefont {Matsukevich}},\ }\href
  {https://journals.aps.org/prl/abstract/10.1103/PhysRevLett.119.150404}
  {\bibfield  {journal} {\bibinfo  {journal} {Phys. Rev. Lett.}\ }\textbf
  {\bibinfo {volume} {119}},\ \bibinfo {pages} {150404} (\bibinfo {year}
  {2017})}\BibitemShut {NoStop}%
\bibitem [{\citenamefont {Zhang}\ \emph {et~al.}(2018)\citenamefont {Zhang},
  \citenamefont {Um}, \citenamefont {Lv}, \citenamefont {Zhang}, \citenamefont
  {Duan},\ and\ \citenamefont {Kim}}]{zhang2018noon}%
  \BibitemOpen
  \bibfield  {author} {\bibinfo {author} {\bibfnamefont {J.}~\bibnamefont
  {Zhang}}, \bibinfo {author} {\bibfnamefont {M.}~\bibnamefont {Um}}, \bibinfo
  {author} {\bibfnamefont {D.}~\bibnamefont {Lv}}, \bibinfo {author}
  {\bibfnamefont {J.-N.}\ \bibnamefont {Zhang}}, \bibinfo {author}
  {\bibfnamefont {L.-M.}\ \bibnamefont {Duan}},\ and\ \bibinfo {author}
  {\bibfnamefont {K.}~\bibnamefont {Kim}},\ }\href
  {https://journals.aps.org/prl/abstract/10.1103/PhysRevLett.121.160502}
  {\bibfield  {journal} {\bibinfo  {journal} {Phys. Rev. Lett.}\ }\textbf
  {\bibinfo {volume} {121}},\ \bibinfo {pages} {160502} (\bibinfo {year}
  {2018})}\BibitemShut {NoStop}%
\bibitem [{\citenamefont {Fl{\"u}hmann}\ \emph {et~al.}(2019)\citenamefont
  {Fl{\"u}hmann}, \citenamefont {Nguyen}, \citenamefont {Marinelli},
  \citenamefont {Negnevitsky}, \citenamefont {Mehta},\ and\ \citenamefont
  {Home}}]{fluhmann2019encoding}%
  \BibitemOpen
  \bibfield  {author} {\bibinfo {author} {\bibfnamefont {C.}~\bibnamefont
  {Fl{\"u}hmann}}, \bibinfo {author} {\bibfnamefont {T.~L.}\ \bibnamefont
  {Nguyen}}, \bibinfo {author} {\bibfnamefont {M.}~\bibnamefont {Marinelli}},
  \bibinfo {author} {\bibfnamefont {V.}~\bibnamefont {Negnevitsky}}, \bibinfo
  {author} {\bibfnamefont {K.}~\bibnamefont {Mehta}},\ and\ \bibinfo {author}
  {\bibfnamefont {J.~P.}\ \bibnamefont {Home}},\ }\href
  {https://www.nature.com/articles/s41586-019-0960-6} {\bibfield  {journal}
  {\bibinfo  {journal} {Nature}\ }\textbf {\bibinfo {volume} {566}},\ \bibinfo
  {pages} {513} (\bibinfo {year} {2019})}\BibitemShut {NoStop}%
\bibitem [{\citenamefont {McCormick}\ \emph {et~al.}(2019)\citenamefont
  {McCormick}, \citenamefont {Keller}, \citenamefont {Burd}, \citenamefont
  {Wineland}, \citenamefont {Wilson},\ and\ \citenamefont
  {Leibfried}}]{mccormick2019quantum}%
  \BibitemOpen
  \bibfield  {author} {\bibinfo {author} {\bibfnamefont {K.~C.}\ \bibnamefont
  {McCormick}}, \bibinfo {author} {\bibfnamefont {J.}~\bibnamefont {Keller}},
  \bibinfo {author} {\bibfnamefont {S.~C.}\ \bibnamefont {Burd}}, \bibinfo
  {author} {\bibfnamefont {D.~J.}\ \bibnamefont {Wineland}}, \bibinfo {author}
  {\bibfnamefont {A.~C.}\ \bibnamefont {Wilson}},\ and\ \bibinfo {author}
  {\bibfnamefont {D.}~\bibnamefont {Leibfried}},\ }\href
  {https://www.nature.com/articles/s41586-019-1421-y} {\bibfield  {journal}
  {\bibinfo  {journal} {Nature}\ }\textbf {\bibinfo {volume} {572}},\ \bibinfo
  {pages} {86} (\bibinfo {year} {2019})}\BibitemShut {NoStop}%
\bibitem [{\citenamefont {Chu}\ \emph {et~al.}(2018)\citenamefont {Chu},
  \citenamefont {Kharel}, \citenamefont {Yoon}, \citenamefont {Frunzio},
  \citenamefont {Rakich},\ and\ \citenamefont {Schoelkopf}}]{chu2018creation}%
  \BibitemOpen
  \bibfield  {author} {\bibinfo {author} {\bibfnamefont {Y.}~\bibnamefont
  {Chu}}, \bibinfo {author} {\bibfnamefont {P.}~\bibnamefont {Kharel}},
  \bibinfo {author} {\bibfnamefont {T.}~\bibnamefont {Yoon}}, \bibinfo {author}
  {\bibfnamefont {L.}~\bibnamefont {Frunzio}}, \bibinfo {author} {\bibfnamefont
  {P.~T.}\ \bibnamefont {Rakich}},\ and\ \bibinfo {author} {\bibfnamefont
  {R.~J.}\ \bibnamefont {Schoelkopf}},\ }\href
  {https://www.nature.com/articles/s41586-018-0717-7} {\bibfield  {journal}
  {\bibinfo  {journal} {Nature}\ }\textbf {\bibinfo {volume} {563}},\ \bibinfo
  {pages} {666} (\bibinfo {year} {2018})}\BibitemShut {NoStop}%
\bibitem [{\citenamefont {Gessner}\ \emph {et~al.}(2019)\citenamefont
  {Gessner}, \citenamefont {Smerzi},\ and\ \citenamefont
  {Pezz{\`e}}}]{gessner2019metrological}%
  \BibitemOpen
  \bibfield  {author} {\bibinfo {author} {\bibfnamefont {M.}~\bibnamefont
  {Gessner}}, \bibinfo {author} {\bibfnamefont {A.}~\bibnamefont {Smerzi}},\
  and\ \bibinfo {author} {\bibfnamefont {L.}~\bibnamefont {Pezz{\`e}}},\ }\href
  {https://journals.aps.org/prl/abstract/10.1103/PhysRevLett.122.090503}
  {\bibfield  {journal} {\bibinfo  {journal} {Phys. Rev. Lett.}\ }\textbf
  {\bibinfo {volume} {122}},\ \bibinfo {pages} {090503} (\bibinfo {year}
  {2019})}\BibitemShut {NoStop}%
\bibitem [{\citenamefont {Wolf}\ \emph {et~al.}(2019)\citenamefont {Wolf},
  \citenamefont {Shi}, \citenamefont {Heip}, \citenamefont {Gessner},
  \citenamefont {Pezz{\`e}}, \citenamefont {Smerzi}, \citenamefont {Schulte},
  \citenamefont {Hammerer},\ and\ \citenamefont {Schmidt}}]{wolf2019motional}%
  \BibitemOpen
  \bibfield  {author} {\bibinfo {author} {\bibfnamefont {F.}~\bibnamefont
  {Wolf}}, \bibinfo {author} {\bibfnamefont {C.}~\bibnamefont {Shi}}, \bibinfo
  {author} {\bibfnamefont {J.~C.}\ \bibnamefont {Heip}}, \bibinfo {author}
  {\bibfnamefont {M.}~\bibnamefont {Gessner}}, \bibinfo {author} {\bibfnamefont
  {L.}~\bibnamefont {Pezz{\`e}}}, \bibinfo {author} {\bibfnamefont
  {A.}~\bibnamefont {Smerzi}}, \bibinfo {author} {\bibfnamefont
  {M.}~\bibnamefont {Schulte}}, \bibinfo {author} {\bibfnamefont
  {K.}~\bibnamefont {Hammerer}},\ and\ \bibinfo {author} {\bibfnamefont
  {P.~O.}\ \bibnamefont {Schmidt}},\ }\href
  {https://www.nature.com/articles/s41467-019-10576-4} {\bibfield  {journal}
  {\bibinfo  {journal} {Nat. Commun.}\ }\textbf {\bibinfo {volume} {10}},\
  \bibinfo {pages} {1} (\bibinfo {year} {2019})}\BibitemShut {NoStop}%
\bibitem [{\citenamefont {Hanamura}\ \emph {et~al.}(2021)\citenamefont
  {Hanamura}, \citenamefont {Asavanant}, \citenamefont {Fukui}, \citenamefont
  {Konno},\ and\ \citenamefont {Furusawa}}]{hanamura2021estimation}%
  \BibitemOpen
  \bibfield  {author} {\bibinfo {author} {\bibfnamefont {F.}~\bibnamefont
  {Hanamura}}, \bibinfo {author} {\bibfnamefont {W.}~\bibnamefont {Asavanant}},
  \bibinfo {author} {\bibfnamefont {K.}~\bibnamefont {Fukui}}, \bibinfo
  {author} {\bibfnamefont {S.}~\bibnamefont {Konno}},\ and\ \bibinfo {author}
  {\bibfnamefont {A.}~\bibnamefont {Furusawa}},\ }\href
  {https://arxiv.org/abs/2102.05276} {\bibfield  {journal} {\bibinfo  {journal}
  {arXiv preprint arXiv:2102.05276}\ } (\bibinfo {year} {2021})}\BibitemShut
  {NoStop}%
\bibitem [{\citenamefont {Duivenvoorden}\ \emph {et~al.}(2017)\citenamefont
  {Duivenvoorden}, \citenamefont {Terhal},\ and\ \citenamefont
  {Weigand}}]{duivenvoorden2017single}%
  \BibitemOpen
  \bibfield  {author} {\bibinfo {author} {\bibfnamefont {K.}~\bibnamefont
  {Duivenvoorden}}, \bibinfo {author} {\bibfnamefont {B.~M.}\ \bibnamefont
  {Terhal}},\ and\ \bibinfo {author} {\bibfnamefont {D.}~\bibnamefont
  {Weigand}},\ }\href
  {https://journals.aps.org/pra/abstract/10.1103/PhysRevA.95.012305} {\bibfield
   {journal} {\bibinfo  {journal} {Physs Rev. A}\ }\textbf {\bibinfo {volume}
  {95}},\ \bibinfo {pages} {012305} (\bibinfo {year} {2017})}\BibitemShut
  {NoStop}%
\bibitem [{\citenamefont {Michael}\ \emph {et~al.}(2016)\citenamefont
  {Michael}, \citenamefont {Silveri}, \citenamefont {Brierley}, \citenamefont
  {Albert}, \citenamefont {Salmilehto}, \citenamefont {Jiang},\ and\
  \citenamefont {Girvin}}]{michael2016new}%
  \BibitemOpen
  \bibfield  {author} {\bibinfo {author} {\bibfnamefont {M.~H.}\ \bibnamefont
  {Michael}}, \bibinfo {author} {\bibfnamefont {M.}~\bibnamefont {Silveri}},
  \bibinfo {author} {\bibfnamefont {R.}~\bibnamefont {Brierley}}, \bibinfo
  {author} {\bibfnamefont {V.~V.}\ \bibnamefont {Albert}}, \bibinfo {author}
  {\bibfnamefont {J.}~\bibnamefont {Salmilehto}}, \bibinfo {author}
  {\bibfnamefont {L.}~\bibnamefont {Jiang}},\ and\ \bibinfo {author}
  {\bibfnamefont {S.~M.}\ \bibnamefont {Girvin}},\ }\href
  {https://journals.aps.org/prx/abstract/10.1103/PhysRevX.6.031006} {\bibfield
  {journal} {\bibinfo  {journal} {Phys. Rev. X}\ }\textbf {\bibinfo {volume}
  {6}},\ \bibinfo {pages} {031006} (\bibinfo {year} {2016})}\BibitemShut
  {NoStop}%
\bibitem [{\citenamefont {Hu}\ \emph {et~al.}(2019)\citenamefont {Hu},
  \citenamefont {Ma}, \citenamefont {Cai}, \citenamefont {Mu}, \citenamefont
  {Xu}, \citenamefont {Wang}, \citenamefont {Wu}, \citenamefont {Wang},
  \citenamefont {Song}, \citenamefont {Zou}, \citenamefont {Girvin},
  \citenamefont {Duan},\ and\ \citenamefont {Sun}}]{hu2019quantum}%
  \BibitemOpen
  \bibfield  {author} {\bibinfo {author} {\bibfnamefont {L.}~\bibnamefont
  {Hu}}, \bibinfo {author} {\bibfnamefont {Y.}~\bibnamefont {Ma}}, \bibinfo
  {author} {\bibfnamefont {W.}~\bibnamefont {Cai}}, \bibinfo {author}
  {\bibfnamefont {X.}~\bibnamefont {Mu}}, \bibinfo {author} {\bibfnamefont
  {Y.}~\bibnamefont {Xu}}, \bibinfo {author} {\bibfnamefont {W.}~\bibnamefont
  {Wang}}, \bibinfo {author} {\bibfnamefont {Y.}~\bibnamefont {Wu}}, \bibinfo
  {author} {\bibfnamefont {H.}~\bibnamefont {Wang}}, \bibinfo {author}
  {\bibfnamefont {Y.}~\bibnamefont {Song}}, \bibinfo {author} {\bibfnamefont
  {C.-L.}\ \bibnamefont {Zou}}, \bibinfo {author} {\bibfnamefont {S.~M.}\
  \bibnamefont {Girvin}}, \bibinfo {author} {\bibfnamefont {L.~M.}\
  \bibnamefont {Duan}},\ and\ \bibinfo {author} {\bibfnamefont
  {L.}~\bibnamefont {Sun}},\ }\href
  {https://www.nature.com/articles/s41567-018-0414-3} {\bibfield  {journal}
  {\bibinfo  {journal} {Nat. Phys.}\ }\textbf {\bibinfo {volume} {15}},\
  \bibinfo {pages} {503} (\bibinfo {year} {2019})}\BibitemShut {NoStop}%
\bibitem [{\citenamefont {Campagne-Ibarcq}\ \emph {et~al.}(2020)\citenamefont
  {Campagne-Ibarcq}, \citenamefont {Eickbusch}, \citenamefont {Touzard},
  \citenamefont {Zalys-Geller}, \citenamefont {Frattini}, \citenamefont
  {Sivak}, \citenamefont {Reinhold}, \citenamefont {Puri}, \citenamefont
  {Shankar}, \citenamefont {Schoelkopf}, \citenamefont {Frunzio}, \citenamefont
  {Mirrahimi},\ and\ \citenamefont {Devoret}}]{campagne2020quantum}%
  \BibitemOpen
  \bibfield  {author} {\bibinfo {author} {\bibfnamefont {P.}~\bibnamefont
  {Campagne-Ibarcq}}, \bibinfo {author} {\bibfnamefont {A.}~\bibnamefont
  {Eickbusch}}, \bibinfo {author} {\bibfnamefont {S.}~\bibnamefont {Touzard}},
  \bibinfo {author} {\bibfnamefont {E.}~\bibnamefont {Zalys-Geller}}, \bibinfo
  {author} {\bibfnamefont {N.~E.}\ \bibnamefont {Frattini}}, \bibinfo {author}
  {\bibfnamefont {V.~V.}\ \bibnamefont {Sivak}}, \bibinfo {author}
  {\bibfnamefont {P.}~\bibnamefont {Reinhold}}, \bibinfo {author}
  {\bibfnamefont {S.}~\bibnamefont {Puri}}, \bibinfo {author} {\bibfnamefont
  {S.}~\bibnamefont {Shankar}}, \bibinfo {author} {\bibfnamefont {R.~J.}\
  \bibnamefont {Schoelkopf}}, \bibinfo {author} {\bibfnamefont
  {L.}~\bibnamefont {Frunzio}}, \bibinfo {author} {\bibfnamefont
  {M.}~\bibnamefont {Mirrahimi}},\ and\ \bibinfo {author} {\bibfnamefont
  {M.~H.}\ \bibnamefont {Devoret}},\ }\href
  {https://www.nature.com/articles/s41586-020-2603-3} {\bibfield  {journal}
  {\bibinfo  {journal} {Nature}\ }\textbf {\bibinfo {volume} {584}},\ \bibinfo
  {pages} {368} (\bibinfo {year} {2020})}\BibitemShut {NoStop}%
\bibitem [{\citenamefont {Ghosh}\ \emph {et~al.}(2018)\citenamefont {Ghosh},
  \citenamefont {Niedenzu}, \citenamefont {Mukherjee},\ and\ \citenamefont
  {Kurizki}}]{ghosh2018thermodynamic}%
  \BibitemOpen
  \bibfield  {author} {\bibinfo {author} {\bibfnamefont {A.}~\bibnamefont
  {Ghosh}}, \bibinfo {author} {\bibfnamefont {W.}~\bibnamefont {Niedenzu}},
  \bibinfo {author} {\bibfnamefont {V.}~\bibnamefont {Mukherjee}},\ and\
  \bibinfo {author} {\bibfnamefont {G.}~\bibnamefont {Kurizki}},\ }in\
  \href@noop {} {\emph {\bibinfo {booktitle} {Thermodynamics in the Quantum
  Regime}}}\ (\bibinfo  {publisher} {Springer},\ \bibinfo {year} {2018})\ pp.\
  \bibinfo {pages} {37--66}\BibitemShut {NoStop}%
\bibitem [{\citenamefont {Maslennikov}\ \emph {et~al.}(2019)\citenamefont
  {Maslennikov}, \citenamefont {Ding}, \citenamefont {Habl{\"u}tzel},
  \citenamefont {Gan}, \citenamefont {Roulet}, \citenamefont {Nimmrichter},
  \citenamefont {Dai}, \citenamefont {Scarani},\ and\ \citenamefont
  {Matsukevich}}]{maslennikov2019quantum}%
  \BibitemOpen
  \bibfield  {author} {\bibinfo {author} {\bibfnamefont {G.}~\bibnamefont
  {Maslennikov}}, \bibinfo {author} {\bibfnamefont {S.}~\bibnamefont {Ding}},
  \bibinfo {author} {\bibfnamefont {R.}~\bibnamefont {Habl{\"u}tzel}}, \bibinfo
  {author} {\bibfnamefont {J.}~\bibnamefont {Gan}}, \bibinfo {author}
  {\bibfnamefont {A.}~\bibnamefont {Roulet}}, \bibinfo {author} {\bibfnamefont
  {S.}~\bibnamefont {Nimmrichter}}, \bibinfo {author} {\bibfnamefont
  {J.}~\bibnamefont {Dai}}, \bibinfo {author} {\bibfnamefont {V.}~\bibnamefont
  {Scarani}},\ and\ \bibinfo {author} {\bibfnamefont {D.}~\bibnamefont
  {Matsukevich}},\ }\href
  {https://www.nature.com/articles/s41467-018-08090-0%C2%A0} {\bibfield
  {journal} {\bibinfo  {journal} {Nat. Commun.}\ }\textbf {\bibinfo {volume}
  {10}},\ \bibinfo {pages} {1} (\bibinfo {year} {2019})}\BibitemShut {NoStop}%
\bibitem [{\citenamefont {Cai}\ \emph {et~al.}(2021)\citenamefont {Cai},
  \citenamefont {Liu}, \citenamefont {Zhao}, \citenamefont {Wu}, \citenamefont
  {Mei}, \citenamefont {Jiang}, \citenamefont {He}, \citenamefont {Zhang},
  \citenamefont {Zhou},\ and\ \citenamefont {Duan}}]{cai2021observation}%
  \BibitemOpen
  \bibfield  {author} {\bibinfo {author} {\bibfnamefont {M.-L.}\ \bibnamefont
  {Cai}}, \bibinfo {author} {\bibfnamefont {Z.-D.}\ \bibnamefont {Liu}},
  \bibinfo {author} {\bibfnamefont {W.-D.}\ \bibnamefont {Zhao}}, \bibinfo
  {author} {\bibfnamefont {Y.-K.}\ \bibnamefont {Wu}}, \bibinfo {author}
  {\bibfnamefont {Q.-X.}\ \bibnamefont {Mei}}, \bibinfo {author} {\bibfnamefont
  {Y.}~\bibnamefont {Jiang}}, \bibinfo {author} {\bibfnamefont
  {L.}~\bibnamefont {He}}, \bibinfo {author} {\bibfnamefont {X.}~\bibnamefont
  {Zhang}}, \bibinfo {author} {\bibfnamefont {Z.-C.}\ \bibnamefont {Zhou}},\
  and\ \bibinfo {author} {\bibfnamefont {L.-M.}\ \bibnamefont {Duan}},\ }\href
  {https://www.nature.com/articles/s41467-021-21425-8} {\bibfield  {journal}
  {\bibinfo  {journal} {Nat. Commun.}\ }\textbf {\bibinfo {volume} {12}},\
  \bibinfo {pages} {1} (\bibinfo {year} {2021})}\BibitemShut {NoStop}%
\bibitem [{\citenamefont {Hudson}(1974)}]{hudson1974wigner}%
  \BibitemOpen
  \bibfield  {author} {\bibinfo {author} {\bibfnamefont {R.~L.}\ \bibnamefont
  {Hudson}},\ }\href {https://doi.org/10.1016/0034-4877(74)90007-X} {\bibfield
  {journal} {\bibinfo  {journal} {Rep. Math. Phys.}\ }\textbf {\bibinfo
  {volume} {6}},\ \bibinfo {pages} {249} (\bibinfo {year} {1974})}\BibitemShut
  {NoStop}%
\bibitem [{\citenamefont {Walschaers}(2021)}]{walschaers2021non}%
  \BibitemOpen
  \bibfield  {author} {\bibinfo {author} {\bibfnamefont {M.}~\bibnamefont
  {Walschaers}},\ }\href
  {https://journals.aps.org/prxquantum/abstract/10.1103/PRXQuantum.2.030204}
  {\bibfield  {journal} {\bibinfo  {journal} {PRX Quantum}\ }\textbf {\bibinfo
  {volume} {2}},\ \bibinfo {pages} {030204} (\bibinfo {year}
  {2021})}\BibitemShut {NoStop}%
\bibitem [{\citenamefont {Schleich}(2011)}]{schleich2011quantum}%
  \BibitemOpen
  \bibfield  {author} {\bibinfo {author} {\bibfnamefont {W.~P.}\ \bibnamefont
  {Schleich}},\ }\href@noop {} {\emph {\bibinfo {title} {Quantum optics in
  phase space}}}\ (\bibinfo  {publisher} {John Wiley \& Sons},\ \bibinfo {year}
  {2011})\BibitemShut {NoStop}%
\bibitem [{\citenamefont {Lachman}\ \emph {et~al.}(2019)\citenamefont
  {Lachman}, \citenamefont {Straka}, \citenamefont {Hlou{\v{s}}ek},
  \citenamefont {Je{\v{z}}ek},\ and\ \citenamefont
  {Filip}}]{lachman2019faithful}%
  \BibitemOpen
  \bibfield  {author} {\bibinfo {author} {\bibfnamefont {L.}~\bibnamefont
  {Lachman}}, \bibinfo {author} {\bibfnamefont {I.}~\bibnamefont {Straka}},
  \bibinfo {author} {\bibfnamefont {J.}~\bibnamefont {Hlou{\v{s}}ek}}, \bibinfo
  {author} {\bibfnamefont {M.}~\bibnamefont {Je{\v{z}}ek}},\ and\ \bibinfo
  {author} {\bibfnamefont {R.}~\bibnamefont {Filip}},\ }\href
  {https://journals.aps.org/prl/abstract/10.1103/PhysRevLett.123.043601}
  {\bibfield  {journal} {\bibinfo  {journal} {Phys. Rev. Lett.}\ }\textbf
  {\bibinfo {volume} {123}},\ \bibinfo {pages} {043601} (\bibinfo {year}
  {2019})}\BibitemShut {NoStop}%
\bibitem [{\citenamefont {Chabaud}\ \emph {et~al.}(2020)\citenamefont
  {Chabaud}, \citenamefont {Markham},\ and\ \citenamefont
  {Grosshans}}]{chabaud2020stellar}%
  \BibitemOpen
  \bibfield  {author} {\bibinfo {author} {\bibfnamefont {U.}~\bibnamefont
  {Chabaud}}, \bibinfo {author} {\bibfnamefont {D.}~\bibnamefont {Markham}},\
  and\ \bibinfo {author} {\bibfnamefont {F.}~\bibnamefont {Grosshans}},\ }\href
  {https://journals.aps.org/prl/abstract/10.1103/PhysRevLett.124.063605}
  {\bibfield  {journal} {\bibinfo  {journal} {Phys. Rev. Lett.}\ }\textbf
  {\bibinfo {volume} {124}},\ \bibinfo {pages} {063605} (\bibinfo {year}
  {2020})}\BibitemShut {NoStop}%
\bibitem [{\citenamefont {Chabaud}\ \emph {et~al.}(2021)\citenamefont
  {Chabaud}, \citenamefont {Roeland}, \citenamefont {Walschaers}, \citenamefont
  {Grosshans}, \citenamefont {Parigi}, \citenamefont {Markham},\ and\
  \citenamefont {Treps}}]{chabaud2021certification}%
  \BibitemOpen
  \bibfield  {author} {\bibinfo {author} {\bibfnamefont {U.}~\bibnamefont
  {Chabaud}}, \bibinfo {author} {\bibfnamefont {G.}~\bibnamefont {Roeland}},
  \bibinfo {author} {\bibfnamefont {M.}~\bibnamefont {Walschaers}}, \bibinfo
  {author} {\bibfnamefont {F.}~\bibnamefont {Grosshans}}, \bibinfo {author}
  {\bibfnamefont {V.}~\bibnamefont {Parigi}}, \bibinfo {author} {\bibfnamefont
  {D.}~\bibnamefont {Markham}},\ and\ \bibinfo {author} {\bibfnamefont
  {N.}~\bibnamefont {Treps}},\ }\href@noop {} {\bibfield  {journal} {\bibinfo
  {journal} {PRX Quantum}\ }\textbf {\bibinfo {volume} {2}},\ \bibinfo {pages}
  {020333} (\bibinfo {year} {2021})}\BibitemShut {NoStop}%
\bibitem [{\citenamefont {Brownnutt}\ \emph {et~al.}(2015)\citenamefont
  {Brownnutt}, \citenamefont {Kumph}, \citenamefont {Rabl},\ and\ \citenamefont
  {Blatt}}]{brownnutt2015ion}%
  \BibitemOpen
  \bibfield  {author} {\bibinfo {author} {\bibfnamefont {M.}~\bibnamefont
  {Brownnutt}}, \bibinfo {author} {\bibfnamefont {M.}~\bibnamefont {Kumph}},
  \bibinfo {author} {\bibfnamefont {P.}~\bibnamefont {Rabl}},\ and\ \bibinfo
  {author} {\bibfnamefont {R.}~\bibnamefont {Blatt}},\ }\href
  {https://journals.aps.org/rmp/pdf/10.1103/RevModPhys.87.1419} {\bibfield
  {journal} {\bibinfo  {journal} {Rev. Mod. Phys.}\ }\textbf {\bibinfo {volume}
  {87}},\ \bibinfo {pages} {1419} (\bibinfo {year} {2015})}\BibitemShut
  {NoStop}%
\bibitem [{\citenamefont {Aspelmeyer}\ \emph {et~al.}(2014)\citenamefont
  {Aspelmeyer}, \citenamefont {Kippenberg},\ and\ \citenamefont
  {Marquardt}}]{aspelmeyer2014cavity}%
  \BibitemOpen
  \bibfield  {author} {\bibinfo {author} {\bibfnamefont {M.}~\bibnamefont
  {Aspelmeyer}}, \bibinfo {author} {\bibfnamefont {T.~J.}\ \bibnamefont
  {Kippenberg}},\ and\ \bibinfo {author} {\bibfnamefont {F.}~\bibnamefont
  {Marquardt}},\ }\href
  {https://journals.aps.org/rmp/abstract/10.1103/RevModPhys.86.1391} {\bibfield
   {journal} {\bibinfo  {journal} {Rev. Mod. Phys.}\ }\textbf {\bibinfo
  {volume} {86}},\ \bibinfo {pages} {1391} (\bibinfo {year}
  {2014})}\BibitemShut {NoStop}%
\bibitem [{\citenamefont {Higginbottom}\ \emph {et~al.}(2016)\citenamefont
  {Higginbottom}, \citenamefont {Slodi{\v{c}}ka}, \citenamefont {Araneda},
  \citenamefont {Lachman}, \citenamefont {Filip}, \citenamefont {Hennrich},\
  and\ \citenamefont {Blatt}}]{higginbottom2016pure}%
  \BibitemOpen
  \bibfield  {author} {\bibinfo {author} {\bibfnamefont {D.~B.}\ \bibnamefont
  {Higginbottom}}, \bibinfo {author} {\bibfnamefont {L.}~\bibnamefont
  {Slodi{\v{c}}ka}}, \bibinfo {author} {\bibfnamefont {G.}~\bibnamefont
  {Araneda}}, \bibinfo {author} {\bibfnamefont {L.}~\bibnamefont {Lachman}},
  \bibinfo {author} {\bibfnamefont {R.}~\bibnamefont {Filip}}, \bibinfo
  {author} {\bibfnamefont {M.}~\bibnamefont {Hennrich}},\ and\ \bibinfo
  {author} {\bibfnamefont {R.}~\bibnamefont {Blatt}},\ }\href
  {http://iopscience.iop.org/article/10.1088/1367-2630/18/9/093038/meta}
  {\bibfield  {journal} {\bibinfo  {journal} {New J. Phys.}\ }\textbf {\bibinfo
  {volume} {18}},\ \bibinfo {pages} {093038} (\bibinfo {year}
  {2016})}\BibitemShut {NoStop}%
\bibitem [{\citenamefont {Straka}\ \emph {et~al.}(2018)\citenamefont {Straka},
  \citenamefont {Lachman}, \citenamefont {Hlou{\v{s}}ek}, \citenamefont
  {Mikov{\'a}}, \citenamefont {Mi{\v{c}}uda}, \citenamefont {Je{\v{z}}ek},\
  and\ \citenamefont {Filip}}]{straka2018quantum}%
  \BibitemOpen
  \bibfield  {author} {\bibinfo {author} {\bibfnamefont {I.}~\bibnamefont
  {Straka}}, \bibinfo {author} {\bibfnamefont {L.}~\bibnamefont {Lachman}},
  \bibinfo {author} {\bibfnamefont {J.}~\bibnamefont {Hlou{\v{s}}ek}}, \bibinfo
  {author} {\bibfnamefont {M.}~\bibnamefont {Mikov{\'a}}}, \bibinfo {author}
  {\bibfnamefont {M.}~\bibnamefont {Mi{\v{c}}uda}}, \bibinfo {author}
  {\bibfnamefont {M.}~\bibnamefont {Je{\v{z}}ek}},\ and\ \bibinfo {author}
  {\bibfnamefont {R.}~\bibnamefont {Filip}},\ }\href
  {https://www.nature.com/articles/s41534-017-0054-y} {\bibfield  {journal}
  {\bibinfo  {journal} {npj Quantum Inf.}\ }\textbf {\bibinfo {volume} {4}},\
  \bibinfo {pages} {4} (\bibinfo {year} {2018})}\BibitemShut {NoStop}%
\bibitem [{\citenamefont {Straka}\ \emph {et~al.}(2014)\citenamefont {Straka},
  \citenamefont {Predojevi{\'c}}, \citenamefont {Huber}, \citenamefont
  {Lachman}, \citenamefont {Butschek}, \citenamefont {Mikov{\'a}},
  \citenamefont {Mi{\v{c}}uda}, \citenamefont {Solomon}, \citenamefont {Weihs},
  \citenamefont {Je{\v{z}}ek},\ and\ \citenamefont
  {Filip}}]{straka2014quantum}%
  \BibitemOpen
  \bibfield  {author} {\bibinfo {author} {\bibfnamefont {I.}~\bibnamefont
  {Straka}}, \bibinfo {author} {\bibfnamefont {A.}~\bibnamefont
  {Predojevi{\'c}}}, \bibinfo {author} {\bibfnamefont {T.}~\bibnamefont
  {Huber}}, \bibinfo {author} {\bibfnamefont {L.}~\bibnamefont {Lachman}},
  \bibinfo {author} {\bibfnamefont {L.}~\bibnamefont {Butschek}}, \bibinfo
  {author} {\bibfnamefont {M.}~\bibnamefont {Mikov{\'a}}}, \bibinfo {author}
  {\bibfnamefont {M.}~\bibnamefont {Mi{\v{c}}uda}}, \bibinfo {author}
  {\bibfnamefont {G.~S.}\ \bibnamefont {Solomon}}, \bibinfo {author}
  {\bibfnamefont {G.}~\bibnamefont {Weihs}}, \bibinfo {author} {\bibfnamefont
  {M.}~\bibnamefont {Je{\v{z}}ek}},\ and\ \bibinfo {author} {\bibfnamefont
  {R.}~\bibnamefont {Filip}},\ }\href
  {https://journals.aps.org/prl/abstract/10.1103/PhysRevLett.113.223603}
  {\bibfield  {journal} {\bibinfo  {journal} {Phys. Rev. Lett.}\ }\textbf
  {\bibinfo {volume} {113}},\ \bibinfo {pages} {223603} (\bibinfo {year}
  {2014})}\BibitemShut {NoStop}%
\bibitem [{\citenamefont {Ra}\ \emph {et~al.}(2020)\citenamefont {Ra},
  \citenamefont {Dufour}, \citenamefont {Walschaers}, \citenamefont {Jacquard},
  \citenamefont {Michel}, \citenamefont {Fabre},\ and\ \citenamefont
  {Treps}}]{ra2020non}%
  \BibitemOpen
  \bibfield  {author} {\bibinfo {author} {\bibfnamefont {Y.-S.}\ \bibnamefont
  {Ra}}, \bibinfo {author} {\bibfnamefont {A.}~\bibnamefont {Dufour}}, \bibinfo
  {author} {\bibfnamefont {M.}~\bibnamefont {Walschaers}}, \bibinfo {author}
  {\bibfnamefont {C.}~\bibnamefont {Jacquard}}, \bibinfo {author}
  {\bibfnamefont {T.}~\bibnamefont {Michel}}, \bibinfo {author} {\bibfnamefont
  {C.}~\bibnamefont {Fabre}},\ and\ \bibinfo {author} {\bibfnamefont
  {N.}~\bibnamefont {Treps}},\ }\href
  {https://www.nature.com/articles/s41567-019-0726-y} {\bibfield  {journal}
  {\bibinfo  {journal} {Nat. Phys.}\ }\textbf {\bibinfo {volume} {16}},\
  \bibinfo {pages} {144} (\bibinfo {year} {2020})}\BibitemShut {NoStop}%
\bibitem [{\citenamefont {Walschaers}\ \emph {et~al.}(2018)\citenamefont
  {Walschaers}, \citenamefont {Sarkar}, \citenamefont {Parigi},\ and\
  \citenamefont {Treps}}]{walschaers2018tailoring}%
  \BibitemOpen
  \bibfield  {author} {\bibinfo {author} {\bibfnamefont {M.}~\bibnamefont
  {Walschaers}}, \bibinfo {author} {\bibfnamefont {S.}~\bibnamefont {Sarkar}},
  \bibinfo {author} {\bibfnamefont {V.}~\bibnamefont {Parigi}},\ and\ \bibinfo
  {author} {\bibfnamefont {N.}~\bibnamefont {Treps}},\ }\href
  {https://journals.aps.org/prl/abstract/10.1103/PhysRevLett.121.220501}
  {\bibfield  {journal} {\bibinfo  {journal} {Phys. Rev. Lett.}\ }\textbf
  {\bibinfo {volume} {121}},\ \bibinfo {pages} {220501} (\bibinfo {year}
  {2018})}\BibitemShut {NoStop}%
\bibitem [{\citenamefont {de~Neeve}\ \emph {et~al.}(2022)\citenamefont
  {de~Neeve}, \citenamefont {Nguyen}, \citenamefont {Behrle},\ and\
  \citenamefont {Home}}]{de2022error}%
  \BibitemOpen
  \bibfield  {author} {\bibinfo {author} {\bibfnamefont {B.}~\bibnamefont
  {de~Neeve}}, \bibinfo {author} {\bibfnamefont {T.-L.}\ \bibnamefont
  {Nguyen}}, \bibinfo {author} {\bibfnamefont {T.}~\bibnamefont {Behrle}},\
  and\ \bibinfo {author} {\bibfnamefont {J.~P.}\ \bibnamefont {Home}},\
  }\href@noop {} {\bibfield  {journal} {\bibinfo  {journal} {Nature Physics}\
  }\textbf {\bibinfo {volume} {18}},\ \bibinfo {pages} {1} (\bibinfo {year}
  {2022})}\BibitemShut {NoStop}%
\bibitem [{\citenamefont {Filip}\ and\ \citenamefont
  {Mi{\v{s}}ta~Jr}(2011)}]{filip2011detecting}%
  \BibitemOpen
  \bibfield  {author} {\bibinfo {author} {\bibfnamefont {R.}~\bibnamefont
  {Filip}}\ and\ \bibinfo {author} {\bibfnamefont {L.}~\bibnamefont
  {Mi{\v{s}}ta~Jr}},\ }\href
  {https://journals.aps.org/prl/abstract/10.1103/PhysRevLett.106.200401}
  {\bibfield  {journal} {\bibinfo  {journal} {Phys. Rev. Lett.}\ }\textbf
  {\bibinfo {volume} {106}},\ \bibinfo {pages} {200401} (\bibinfo {year}
  {2011})}\BibitemShut {NoStop}%
\bibitem [{\citenamefont {Zapletal}\ \emph {et~al.}(2021)\citenamefont
  {Zapletal}, \citenamefont {Darras}, \citenamefont {Le~Jeannic}, \citenamefont
  {Cavaill{\`e}s}, \citenamefont {Guccione}, \citenamefont {Laurat},\ and\
  \citenamefont {Filip}}]{zapletal2021experimental}%
  \BibitemOpen
  \bibfield  {author} {\bibinfo {author} {\bibfnamefont {P.}~\bibnamefont
  {Zapletal}}, \bibinfo {author} {\bibfnamefont {T.}~\bibnamefont {Darras}},
  \bibinfo {author} {\bibfnamefont {H.}~\bibnamefont {Le~Jeannic}}, \bibinfo
  {author} {\bibfnamefont {A.}~\bibnamefont {Cavaill{\`e}s}}, \bibinfo {author}
  {\bibfnamefont {G.}~\bibnamefont {Guccione}}, \bibinfo {author}
  {\bibfnamefont {J.}~\bibnamefont {Laurat}},\ and\ \bibinfo {author}
  {\bibfnamefont {R.}~\bibnamefont {Filip}},\ }\href
  {https://www.osapublishing.org/optica/fulltext.cfm?uri=optica-8-5-743&id=451164}
  {\bibfield  {journal} {\bibinfo  {journal} {Optica}\ }\textbf {\bibinfo
  {volume} {8}},\ \bibinfo {pages} {743} (\bibinfo {year} {2021})}\BibitemShut
  {NoStop}%
\bibitem [{\citenamefont {Gan}\ \emph {et~al.}(2020)\citenamefont {Gan},
  \citenamefont {Maslennikov}, \citenamefont {Tseng}, \citenamefont {Nguyen},\
  and\ \citenamefont {Matsukevich}}]{gan2020hybrid}%
  \BibitemOpen
  \bibfield  {author} {\bibinfo {author} {\bibfnamefont {H.}~\bibnamefont
  {Gan}}, \bibinfo {author} {\bibfnamefont {G.}~\bibnamefont {Maslennikov}},
  \bibinfo {author} {\bibfnamefont {K.-W.}\ \bibnamefont {Tseng}}, \bibinfo
  {author} {\bibfnamefont {C.}~\bibnamefont {Nguyen}},\ and\ \bibinfo {author}
  {\bibfnamefont {D.}~\bibnamefont {Matsukevich}},\ }\href
  {https://journals.aps.org/prl/abstract/10.1103/PhysRevLett.124.170502}
  {\bibfield  {journal} {\bibinfo  {journal} {Phys. Rev. Lett.}\ }\textbf
  {\bibinfo {volume} {124}},\ \bibinfo {pages} {170502} (\bibinfo {year}
  {2020})}\BibitemShut {NoStop}%
\bibitem [{S2()}]{S2}%
  \BibitemOpen
  \href@noop {} {\bibinfo  {journal} {See the Supplemental Material S2 for
  experimental details of preparation, measurement, and evaluation of
  phonon-number distributions of presented motional states. It includes the
  references~\cite{cirac1993preparationS, eschner1995stochasticS,
  cirac1994nonclassicalS,blatt1995trappingS,myatt2000decoherenceS,podhora2020unconditionalS,um2016phononS,kienzler2015quantumS,schneider1998decoherenceS,murao1998decoherence,bonifacio2000model,budini2002localization}}\
  }\BibitemShut {NoStop}%
\bibitem [{\citenamefont {Leibfried}\ \emph {et~al.}(2003)\citenamefont
  {Leibfried}, \citenamefont {Blatt}, \citenamefont {Monroe},\ and\
  \citenamefont {Wineland}}]{leibfried2003quantum}%
  \BibitemOpen
\bibfield  {journal} {  }\bibfield  {author} {\bibinfo {author} {\bibfnamefont
  {D.}~\bibnamefont {Leibfried}}, \bibinfo {author} {\bibfnamefont
  {R.}~\bibnamefont {Blatt}}, \bibinfo {author} {\bibfnamefont
  {C.}~\bibnamefont {Monroe}},\ and\ \bibinfo {author} {\bibfnamefont
  {D.}~\bibnamefont {Wineland}},\ }\href
  {https://journals.aps.org/rmp/abstract/10.1103/RevModPhys.75.281} {\bibfield
  {journal} {\bibinfo  {journal} {Rev. Mod. Phys.}\ }\textbf {\bibinfo {volume}
  {75}},\ \bibinfo {pages} {281} (\bibinfo {year} {2003})}\BibitemShut
  {NoStop}%
\bibitem [{\citenamefont {Eschner}\ \emph {et~al.}(2003)\citenamefont
  {Eschner}, \citenamefont {Morigi}, \citenamefont {Schmidt-Kaler},\ and\
  \citenamefont {Blatt}}]{eschner2003laser}%
  \BibitemOpen
  \bibfield  {author} {\bibinfo {author} {\bibfnamefont {J.}~\bibnamefont
  {Eschner}}, \bibinfo {author} {\bibfnamefont {G.}~\bibnamefont {Morigi}},
  \bibinfo {author} {\bibfnamefont {F.}~\bibnamefont {Schmidt-Kaler}},\ and\
  \bibinfo {author} {\bibfnamefont {R.}~\bibnamefont {Blatt}},\ }\href
  {https://www.osapublishing.org/josab/abstract.cfm?uri=JOSAB-20-5-1003}
  {\bibfield  {journal} {\bibinfo  {journal} {J. Opt. Soc. Am. B}\ }\textbf
  {\bibinfo {volume} {20}},\ \bibinfo {pages} {1003} (\bibinfo {year}
  {2003})}\BibitemShut {NoStop}%
\bibitem [{\citenamefont {Hempel}\ \emph {et~al.}(2013)\citenamefont {Hempel},
  \citenamefont {Lanyon}, \citenamefont {Jurcevic}, \citenamefont {Gerritsma},
  \citenamefont {Blatt},\ and\ \citenamefont {Roos}}]{hempel2013entanglement}%
  \BibitemOpen
  \bibfield  {author} {\bibinfo {author} {\bibfnamefont {C.}~\bibnamefont
  {Hempel}}, \bibinfo {author} {\bibfnamefont {B.}~\bibnamefont {Lanyon}},
  \bibinfo {author} {\bibfnamefont {P.}~\bibnamefont {Jurcevic}}, \bibinfo
  {author} {\bibfnamefont {R.}~\bibnamefont {Gerritsma}}, \bibinfo {author}
  {\bibfnamefont {R.}~\bibnamefont {Blatt}},\ and\ \bibinfo {author}
  {\bibfnamefont {C.}~\bibnamefont {Roos}},\ }\href
  {https://www.nature.com/articles/nphoton.2013.172} {\bibfield  {journal}
  {\bibinfo  {journal} {Nat. Photonics}\ }\textbf {\bibinfo {volume} {7}},\
  \bibinfo {pages} {630} (\bibinfo {year} {2013})}\BibitemShut {NoStop}%
\bibitem [{\citenamefont {Wan}\ \emph {et~al.}(2014)\citenamefont {Wan},
  \citenamefont {Gebert}, \citenamefont {W{\"u}bbena}, \citenamefont
  {Scharnhorst}, \citenamefont {Amairi}, \citenamefont {Leroux}, \citenamefont
  {Hemmerling}, \citenamefont {L{\"o}rch}, \citenamefont {Hammerer},\ and\
  \citenamefont {Schmidt}}]{wan2014precision}%
  \BibitemOpen
  \bibfield  {author} {\bibinfo {author} {\bibfnamefont {Y.}~\bibnamefont
  {Wan}}, \bibinfo {author} {\bibfnamefont {F.}~\bibnamefont {Gebert}},
  \bibinfo {author} {\bibfnamefont {J.~B.}\ \bibnamefont {W{\"u}bbena}},
  \bibinfo {author} {\bibfnamefont {N.}~\bibnamefont {Scharnhorst}}, \bibinfo
  {author} {\bibfnamefont {S.}~\bibnamefont {Amairi}}, \bibinfo {author}
  {\bibfnamefont {I.~D.}\ \bibnamefont {Leroux}}, \bibinfo {author}
  {\bibfnamefont {B.}~\bibnamefont {Hemmerling}}, \bibinfo {author}
  {\bibfnamefont {N.}~\bibnamefont {L{\"o}rch}}, \bibinfo {author}
  {\bibfnamefont {K.}~\bibnamefont {Hammerer}},\ and\ \bibinfo {author}
  {\bibfnamefont {P.~O.}\ \bibnamefont {Schmidt}},\ }\href
  {https://www.nature.com/articles/ncomms4096} {\bibfield  {journal} {\bibinfo
  {journal} {Nat. Commun.}\ }\textbf {\bibinfo {volume} {5}},\ \bibinfo {pages}
  {1} (\bibinfo {year} {2014})}\BibitemShut {NoStop}%
\bibitem [{\citenamefont {Gely}\ \emph {et~al.}(2019)\citenamefont {Gely},
  \citenamefont {Kounalakis}, \citenamefont {Dickel}, \citenamefont {Dalle},
  \citenamefont {Vatr{\'e}}, \citenamefont {Baker}, \citenamefont {Jenkins},\
  and\ \citenamefont {Steele}}]{gely2019observation}%
  \BibitemOpen
  \bibfield  {author} {\bibinfo {author} {\bibfnamefont {M.~F.}\ \bibnamefont
  {Gely}}, \bibinfo {author} {\bibfnamefont {M.}~\bibnamefont {Kounalakis}},
  \bibinfo {author} {\bibfnamefont {C.}~\bibnamefont {Dickel}}, \bibinfo
  {author} {\bibfnamefont {J.}~\bibnamefont {Dalle}}, \bibinfo {author}
  {\bibfnamefont {R.}~\bibnamefont {Vatr{\'e}}}, \bibinfo {author}
  {\bibfnamefont {B.}~\bibnamefont {Baker}}, \bibinfo {author} {\bibfnamefont
  {M.~D.}\ \bibnamefont {Jenkins}},\ and\ \bibinfo {author} {\bibfnamefont
  {G.~A.}\ \bibnamefont {Steele}},\ }\href
  {https://www.science.org/doi/abs/10.1126/science.aaw3101} {\bibfield
  {journal} {\bibinfo  {journal} {Science}\ }\textbf {\bibinfo {volume}
  {363}},\ \bibinfo {pages} {1072} (\bibinfo {year} {2019})}\BibitemShut
  {NoStop}%
\bibitem [{\citenamefont {Gelbwaser-Klimovsky}\ and\ \citenamefont
  {Kurizki}(2015)}]{gelbwaser2015work}%
  \BibitemOpen
  \bibfield  {author} {\bibinfo {author} {\bibfnamefont {D.}~\bibnamefont
  {Gelbwaser-Klimovsky}}\ and\ \bibinfo {author} {\bibfnamefont
  {G.}~\bibnamefont {Kurizki}},\ }\href
  {https://www.nature.com/articles/srep07809} {\bibfield  {journal} {\bibinfo
  {journal} {Sci. Rep.}\ }\textbf {\bibinfo {volume} {5}},\ \bibinfo {pages}
  {1} (\bibinfo {year} {2015})}\BibitemShut {NoStop}%
\bibitem [{\citenamefont {Tse}\ \emph {et~al.}(2019)\citenamefont {Tse} \emph
  {et~al.}}]{tse2019quantum}%
  \BibitemOpen
  \bibfield  {author} {\bibinfo {author} {\bibfnamefont {M.}~\bibnamefont
  {Tse}} \emph {et~al.},\ }\href
  {https://journals.aps.org/prl/abstract/10.1103/PhysRevLett.123.231107}
  {\bibfield  {journal} {\bibinfo  {journal} {Phys. Rev. Lett.}\ }\textbf
  {\bibinfo {volume} {123}},\ \bibinfo {pages} {231107} (\bibinfo {year}
  {2019})}\BibitemShut {NoStop}%
\bibitem [{\citenamefont {Gorecki}\ \emph {et~al.}(2020)\citenamefont
  {Gorecki}, \citenamefont {Zhou}, \citenamefont {Jiang},\ and\ \citenamefont
  {Demkowicz-Dobrza{\'n}ski}}]{gorecki2020optimal}%
  \BibitemOpen
  \bibfield  {author} {\bibinfo {author} {\bibfnamefont {W.}~\bibnamefont
  {Gorecki}}, \bibinfo {author} {\bibfnamefont {S.}~\bibnamefont {Zhou}},
  \bibinfo {author} {\bibfnamefont {L.}~\bibnamefont {Jiang}},\ and\ \bibinfo
  {author} {\bibfnamefont {R.}~\bibnamefont {Demkowicz-Dobrza{\'n}ski}},\
  }\href {https://quantum-journal.org/papers/q-2020-07-02-288/} {\bibfield
  {journal} {\bibinfo  {journal} {Quantum}\ }\textbf {\bibinfo {volume} {4}},\
  \bibinfo {pages} {288} (\bibinfo {year} {2020})}\BibitemShut {NoStop}%
\bibitem [{\citenamefont {Kubica}\ and\ \citenamefont
  {Demkowicz-Dobrza{\'n}ski}(2021)}]{kubica2021using}%
  \BibitemOpen
  \bibfield  {author} {\bibinfo {author} {\bibfnamefont {A.}~\bibnamefont
  {Kubica}}\ and\ \bibinfo {author} {\bibfnamefont {R.}~\bibnamefont
  {Demkowicz-Dobrza{\'n}ski}},\ }\href
  {https://journals.aps.org/prl/abstract/10.1103/PhysRevLett.126.150503}
  {\bibfield  {journal} {\bibinfo  {journal} {Phys. Rev. Lett.}\ }\textbf
  {\bibinfo {volume} {126}},\ \bibinfo {pages} {150503} (\bibinfo {year}
  {2021})}\BibitemShut {NoStop}%
\bibitem [{\citenamefont {Schreppler}\ \emph {et~al.}(2014)\citenamefont
  {Schreppler}, \citenamefont {Spethmann}, \citenamefont {Brahms},
  \citenamefont {Botter}, \citenamefont {Barrios},\ and\ \citenamefont
  {Stamper-Kurn}}]{schreppler2014optically}%
  \BibitemOpen
  \bibfield  {author} {\bibinfo {author} {\bibfnamefont {S.}~\bibnamefont
  {Schreppler}}, \bibinfo {author} {\bibfnamefont {N.}~\bibnamefont
  {Spethmann}}, \bibinfo {author} {\bibfnamefont {N.}~\bibnamefont {Brahms}},
  \bibinfo {author} {\bibfnamefont {T.}~\bibnamefont {Botter}}, \bibinfo
  {author} {\bibfnamefont {M.}~\bibnamefont {Barrios}},\ and\ \bibinfo {author}
  {\bibfnamefont {D.~M.}\ \bibnamefont {Stamper-Kurn}},\ }\href
  {https://www.science.org/doi/abs/10.1126/science.1249850} {\bibfield
  {journal} {\bibinfo  {journal} {Science}\ }\textbf {\bibinfo {volume}
  {344}},\ \bibinfo {pages} {1486} (\bibinfo {year} {2014})}\BibitemShut
  {NoStop}%
\bibitem [{\citenamefont {Hebestreit}\ \emph {et~al.}(2018)\citenamefont
  {Hebestreit}, \citenamefont {Frimmer}, \citenamefont {Reimann},\ and\
  \citenamefont {Novotny}}]{hebestreit2018sensing}%
  \BibitemOpen
  \bibfield  {author} {\bibinfo {author} {\bibfnamefont {E.}~\bibnamefont
  {Hebestreit}}, \bibinfo {author} {\bibfnamefont {M.}~\bibnamefont {Frimmer}},
  \bibinfo {author} {\bibfnamefont {R.}~\bibnamefont {Reimann}},\ and\ \bibinfo
  {author} {\bibfnamefont {L.}~\bibnamefont {Novotny}},\ }\href
  {https://journals.aps.org/prl/abstract/10.1103/PhysRevLett.121.063602}
  {\bibfield  {journal} {\bibinfo  {journal} {Phys. Rev. Lett.}\ }\textbf
  {\bibinfo {volume} {121}},\ \bibinfo {pages} {063602} (\bibinfo {year}
  {2018})}\BibitemShut {NoStop}%
\bibitem [{\citenamefont {Cirac}\ \emph {et~al.}(1993)\citenamefont {Cirac},
  \citenamefont {Blatt}, \citenamefont {Parkins},\ and\ \citenamefont
  {Zoller}}]{cirac1993preparationS}%
  \BibitemOpen
  \bibfield  {author} {\bibinfo {author} {\bibfnamefont {J.~I.}\ \bibnamefont
  {Cirac}}, \bibinfo {author} {\bibfnamefont {R.}~\bibnamefont {Blatt}},
  \bibinfo {author} {\bibfnamefont {A.}~\bibnamefont {Parkins}},\ and\ \bibinfo
  {author} {\bibfnamefont {P.}~\bibnamefont {Zoller}},\ }\href
  {https://journals.aps.org/prl/abstract/10.1103/PhysRevLett.70.762} {\bibfield
   {journal} {\bibinfo  {journal} {Phys. Rev. Lett.}\ }\textbf {\bibinfo
  {volume} {70}},\ \bibinfo {pages} {762} (\bibinfo {year} {1993})}\BibitemShut
  {NoStop}%
\bibitem [{\citenamefont {Eschner}\ \emph {et~al.}(1995)\citenamefont
  {Eschner}, \citenamefont {Appasamy},\ and\ \citenamefont
  {Toschek}}]{eschner1995stochasticS}%
  \BibitemOpen
  \bibfield  {author} {\bibinfo {author} {\bibfnamefont {J.}~\bibnamefont
  {Eschner}}, \bibinfo {author} {\bibfnamefont {B.}~\bibnamefont {Appasamy}},\
  and\ \bibinfo {author} {\bibfnamefont {P.}~\bibnamefont {Toschek}},\ }\href
  {https://journals.aps.org/prl/abstract/10.1103/PhysRevLett.74.2435}
  {\bibfield  {journal} {\bibinfo  {journal} {Phys. Rev. Lett.}\ }\textbf
  {\bibinfo {volume} {74}},\ \bibinfo {pages} {2435} (\bibinfo {year}
  {1995})}\BibitemShut {NoStop}%
\bibitem [{\citenamefont {Cirac}\ \emph {et~al.}(1994)\citenamefont {Cirac},
  \citenamefont {Blatt},\ and\ \citenamefont
  {Zoller}}]{cirac1994nonclassicalS}%
  \BibitemOpen
  \bibfield  {author} {\bibinfo {author} {\bibfnamefont {J.~I.}\ \bibnamefont
  {Cirac}}, \bibinfo {author} {\bibfnamefont {R.}~\bibnamefont {Blatt}},\ and\
  \bibinfo {author} {\bibfnamefont {P.}~\bibnamefont {Zoller}},\ }\href
  {https://journals.aps.org/pra/abstract/10.1103/PhysRevA.49.R3174} {\bibfield
  {journal} {\bibinfo  {journal} {Phys. Rev. A}\ }\textbf {\bibinfo {volume}
  {49}},\ \bibinfo {pages} {R3174} (\bibinfo {year} {1994})}\BibitemShut
  {NoStop}%
\bibitem [{\citenamefont {Blatt}\ \emph {et~al.}(1995)\citenamefont {Blatt},
  \citenamefont {Cirac},\ and\ \citenamefont {Zoller}}]{blatt1995trappingS}%
  \BibitemOpen
  \bibfield  {author} {\bibinfo {author} {\bibfnamefont {R.}~\bibnamefont
  {Blatt}}, \bibinfo {author} {\bibfnamefont {J.~I.}\ \bibnamefont {Cirac}},\
  and\ \bibinfo {author} {\bibfnamefont {P.}~\bibnamefont {Zoller}},\ }\href
  {https://journals.aps.org/pra/abstract/10.1103/PhysRevA.52.518} {\bibfield
  {journal} {\bibinfo  {journal} {Phys. Rev. A}\ }\textbf {\bibinfo {volume}
  {52}},\ \bibinfo {pages} {518} (\bibinfo {year} {1995})}\BibitemShut
  {NoStop}%
\bibitem [{\citenamefont {Myatt}\ \emph {et~al.}(2000)\citenamefont {Myatt},
  \citenamefont {King}, \citenamefont {Turchette}, \citenamefont {Sackett},
  \citenamefont {Kielpinski}, \citenamefont {Itano}, \citenamefont {Monroe},\
  and\ \citenamefont {Wineland}}]{myatt2000decoherenceS}%
  \BibitemOpen
  \bibfield  {author} {\bibinfo {author} {\bibfnamefont {C.~J.}\ \bibnamefont
  {Myatt}}, \bibinfo {author} {\bibfnamefont {B.~E.}\ \bibnamefont {King}},
  \bibinfo {author} {\bibfnamefont {Q.~A.}\ \bibnamefont {Turchette}}, \bibinfo
  {author} {\bibfnamefont {C.~A.}\ \bibnamefont {Sackett}}, \bibinfo {author}
  {\bibfnamefont {D.}~\bibnamefont {Kielpinski}}, \bibinfo {author}
  {\bibfnamefont {W.~M.}\ \bibnamefont {Itano}}, \bibinfo {author}
  {\bibfnamefont {C.}~\bibnamefont {Monroe}},\ and\ \bibinfo {author}
  {\bibfnamefont {D.~J.}\ \bibnamefont {Wineland}},\ }\href
  {https://www.nature.com/articles/35002001} {\bibfield  {journal} {\bibinfo
  {journal} {Nature}\ }\textbf {\bibinfo {volume} {403}},\ \bibinfo {pages}
  {269} (\bibinfo {year} {2000})}\BibitemShut {NoStop}%
\bibitem [{\citenamefont {Podhora}\ \emph {et~al.}(2020)\citenamefont
  {Podhora}, \citenamefont {Pham}, \citenamefont {Le{\v{s}}und{\'a}k},
  \citenamefont {Ob{\v{s}}il}, \citenamefont {{\v{C}}{\'\i}{\v{z}}ek},
  \citenamefont {{\v{C}}{\'\i}p}, \citenamefont {Marek}, \citenamefont
  {Slodi{\v{c}}ka},\ and\ \citenamefont {Filip}}]{podhora2020unconditionalS}%
  \BibitemOpen
  \bibfield  {author} {\bibinfo {author} {\bibfnamefont {L.}~\bibnamefont
  {Podhora}}, \bibinfo {author} {\bibfnamefont {T.}~\bibnamefont {Pham}},
  \bibinfo {author} {\bibfnamefont {A.}~\bibnamefont {Le{\v{s}}und{\'a}k}},
  \bibinfo {author} {\bibfnamefont {P.}~\bibnamefont {Ob{\v{s}}il}}, \bibinfo
  {author} {\bibfnamefont {M.}~\bibnamefont {{\v{C}}{\'\i}{\v{z}}ek}}, \bibinfo
  {author} {\bibfnamefont {O.}~\bibnamefont {{\v{C}}{\'\i}p}}, \bibinfo
  {author} {\bibfnamefont {P.}~\bibnamefont {Marek}}, \bibinfo {author}
  {\bibfnamefont {L.}~\bibnamefont {Slodi{\v{c}}ka}},\ and\ \bibinfo {author}
  {\bibfnamefont {R.}~\bibnamefont {Filip}},\ }\href
  {https://onlinelibrary.wiley.com/doi/abs/10.1002/qute.202000012} {\bibfield
  {journal} {\bibinfo  {journal} {Adv. Quantum Technol.}\ }\textbf {\bibinfo
  {volume} {3}},\ \bibinfo {pages} {2000012} (\bibinfo {year}
  {2020})}\BibitemShut {NoStop}%
\bibitem [{\citenamefont {Um}\ \emph {et~al.}(2016)\citenamefont {Um},
  \citenamefont {Zhang}, \citenamefont {Lv}, \citenamefont {Lu}, \citenamefont
  {An}, \citenamefont {Zhang}, \citenamefont {Nha}, \citenamefont {Kim},\ and\
  \citenamefont {Kim}}]{um2016phononS}%
  \BibitemOpen
  \bibfield  {author} {\bibinfo {author} {\bibfnamefont {M.}~\bibnamefont
  {Um}}, \bibinfo {author} {\bibfnamefont {J.}~\bibnamefont {Zhang}}, \bibinfo
  {author} {\bibfnamefont {D.}~\bibnamefont {Lv}}, \bibinfo {author}
  {\bibfnamefont {Y.}~\bibnamefont {Lu}}, \bibinfo {author} {\bibfnamefont
  {S.}~\bibnamefont {An}}, \bibinfo {author} {\bibfnamefont {J.-N.}\
  \bibnamefont {Zhang}}, \bibinfo {author} {\bibfnamefont {H.}~\bibnamefont
  {Nha}}, \bibinfo {author} {\bibfnamefont {M.~S.}\ \bibnamefont {Kim}},\ and\
  \bibinfo {author} {\bibfnamefont {K.}~\bibnamefont {Kim}},\ }\href
  {https://www.nature.com/articles/ncomms11410} {\bibfield  {journal} {\bibinfo
   {journal} {Nat. Commun.}\ }\textbf {\bibinfo {volume} {7}},\ \bibinfo
  {pages} {11410} (\bibinfo {year} {2016})}\BibitemShut {NoStop}%
\bibitem [{\citenamefont {Kienzler}\ \emph {et~al.}(2015)\citenamefont
  {Kienzler}, \citenamefont {Lo}, \citenamefont {Keitch}, \citenamefont
  {De~Clercq}, \citenamefont {Leupold}, \citenamefont {Lindenfelser},
  \citenamefont {Marinelli}, \citenamefont {Negnevitsky},\ and\ \citenamefont
  {Home}}]{kienzler2015quantumS}%
  \BibitemOpen
  \bibfield  {author} {\bibinfo {author} {\bibfnamefont {D.}~\bibnamefont
  {Kienzler}}, \bibinfo {author} {\bibfnamefont {H.-Y.}\ \bibnamefont {Lo}},
  \bibinfo {author} {\bibfnamefont {B.}~\bibnamefont {Keitch}}, \bibinfo
  {author} {\bibfnamefont {L.}~\bibnamefont {De~Clercq}}, \bibinfo {author}
  {\bibfnamefont {F.}~\bibnamefont {Leupold}}, \bibinfo {author} {\bibfnamefont
  {F.}~\bibnamefont {Lindenfelser}}, \bibinfo {author} {\bibfnamefont
  {M.}~\bibnamefont {Marinelli}}, \bibinfo {author} {\bibfnamefont
  {V.}~\bibnamefont {Negnevitsky}},\ and\ \bibinfo {author} {\bibfnamefont
  {J.~P.}\ \bibnamefont {Home}},\ }\href
  {https://www.science.org/doi/abs/10.1126/science.1261033} {\bibfield
  {journal} {\bibinfo  {journal} {Science}\ }\textbf {\bibinfo {volume}
  {347}},\ \bibinfo {pages} {53} (\bibinfo {year} {2015})}\BibitemShut
  {NoStop}%
\bibitem [{\citenamefont {Schneider}\ and\ \citenamefont
  {Milburn}(1998)}]{schneider1998decoherenceS}%
  \BibitemOpen
  \bibfield  {author} {\bibinfo {author} {\bibfnamefont {S.}~\bibnamefont
  {Schneider}}\ and\ \bibinfo {author} {\bibfnamefont {G.~J.}\ \bibnamefont
  {Milburn}},\ }\href
  {https://journals.aps.org/pra/abstract/10.1103/PhysRevA.57.3748} {\bibfield
  {journal} {\bibinfo  {journal} {Phys. Rev. A}\ }\textbf {\bibinfo {volume}
  {57}},\ \bibinfo {pages} {3748} (\bibinfo {year} {1998})}\BibitemShut
  {NoStop}%
\bibitem [{\citenamefont {Murao}\ and\ \citenamefont
  {Knight}(1998)}]{murao1998decoherence}%
  \BibitemOpen
  \bibfield  {author} {\bibinfo {author} {\bibfnamefont {M.}~\bibnamefont
  {Murao}}\ and\ \bibinfo {author} {\bibfnamefont {P.}~\bibnamefont {Knight}},\
  }\href {https://journals.aps.org/pra/abstract/10.1103/PhysRevA.58.663}
  {\bibfield  {journal} {\bibinfo  {journal} {Phys. Rev. A}\ }\textbf {\bibinfo
  {volume} {58}},\ \bibinfo {pages} {663} (\bibinfo {year} {1998})}\BibitemShut
  {NoStop}%
\bibitem [{\citenamefont {Bonifacio}\ \emph {et~al.}(2000)\citenamefont
  {Bonifacio}, \citenamefont {Olivares}, \citenamefont {Tombesi},\ and\
  \citenamefont {Vitali}}]{bonifacio2000model}%
  \BibitemOpen
  \bibfield  {author} {\bibinfo {author} {\bibfnamefont {R.}~\bibnamefont
  {Bonifacio}}, \bibinfo {author} {\bibfnamefont {S.}~\bibnamefont {Olivares}},
  \bibinfo {author} {\bibfnamefont {P.}~\bibnamefont {Tombesi}},\ and\ \bibinfo
  {author} {\bibfnamefont {D.}~\bibnamefont {Vitali}},\ }\href
  {https://journals.aps.org/pra/abstract/10.1103/PhysRevA.61.053802} {\bibfield
   {journal} {\bibinfo  {journal} {Phys. Rev. A}\ }\textbf {\bibinfo {volume}
  {61}},\ \bibinfo {pages} {053802} (\bibinfo {year} {2000})}\BibitemShut
  {NoStop}%
\bibitem [{\citenamefont {Budini}\ \emph {et~al.}(2002)\citenamefont {Budini},
  \citenamefont {de~Matos~Filho},\ and\ \citenamefont
  {Zagury}}]{budini2002localization}%
  \BibitemOpen
  \bibfield  {author} {\bibinfo {author} {\bibfnamefont {A.}~\bibnamefont
  {Budini}}, \bibinfo {author} {\bibfnamefont {R.}~\bibnamefont
  {de~Matos~Filho}},\ and\ \bibinfo {author} {\bibfnamefont {N.}~\bibnamefont
  {Zagury}},\ }\href
  {https://journals.aps.org/pra/abstract/10.1103/PhysRevA.65.041402} {\bibfield
   {journal} {\bibinfo  {journal} {Phys. Rev. A}\ }\textbf {\bibinfo {volume}
  {65}},\ \bibinfo {pages} {041402} (\bibinfo {year} {2002})}\BibitemShut
  {NoStop}%
\end{thebibliography}

\begin{thebibliography}{23}%
\makeatletter
\providecommand \@ifxundefined [1]{%
 \@ifx{#1\undefined}
}%
\providecommand \@ifnum [1]{%
 \ifnum #1\expandafter \@firstoftwo
 \else \expandafter \@secondoftwo
 \fi
}%
\providecommand \@ifx [1]{%
 \ifx #1\expandafter \@firstoftwo
 \else \expandafter \@secondoftwo
 \fi
}%
\providecommand \natexlab [1]{#1}%
\providecommand \enquote  [1]{``#1''}%
\providecommand \bibnamefont  [1]{#1}%
\providecommand \bibfnamefont [1]{#1}%
\providecommand \citenamefont [1]{#1}%
\providecommand \href@noop [0]{\@secondoftwo}%
\providecommand \href [0]{\begingroup \@sanitize@url \@href}%
\providecommand \@href[1]{\@@startlink{#1}\@@href}%
\providecommand \@@href[1]{\endgroup#1\@@endlink}%
\providecommand \@sanitize@url [0]{\catcode `\\12\catcode `\$12\catcode
  `\&12\catcode `\#12\catcode `\^12\catcode `\_12\catcode `\%12\relax}%
\providecommand \@@startlink[1]{}%
\providecommand \@@endlink[0]{}%
\providecommand \url  [0]{\begingroup\@sanitize@url \@url }%
\providecommand \@url [1]{\endgroup\@href {#1}{\urlprefix }}%
\providecommand \urlprefix  [0]{URL }%
\providecommand \Eprint [0]{\href }%
\providecommand \doibase [0]{https://doi.org/}%
\providecommand \selectlanguage [0]{\@gobble}%
\providecommand \bibinfo  [0]{\@secondoftwo}%
\providecommand \bibfield  [0]{\@secondoftwo}%
\providecommand \translation [1]{[#1]}%
\providecommand \BibitemOpen [0]{}%
\providecommand \bibitemStop [0]{}%
\providecommand \bibitemNoStop [0]{.\EOS\space}%
\providecommand \EOS [0]{\spacefactor3000\relax}%
\providecommand \BibitemShut  [1]{\csname bibitem#1\endcsname}%
\let\auto@bib@innerbib\@empty
\bibitem [{\citenamefont {Lachman}\ \emph {et~al.}(2019)\citenamefont
  {Lachman}, \citenamefont {Straka}, \citenamefont {Hlou{\v{s}}ek},
  \citenamefont {Je{\v{z}}ek},\ and\ \citenamefont
  {Filip}}]{lachman2019faithfulS}%
  \BibitemOpen
  \bibfield  {author} {\bibinfo {author} {\bibfnamefont {L.}~\bibnamefont
  {Lachman}}, \bibinfo {author} {\bibfnamefont {I.}~\bibnamefont {Straka}},
  \bibinfo {author} {\bibfnamefont {J.}~\bibnamefont {Hlou{\v{s}}ek}}, \bibinfo
  {author} {\bibfnamefont {M.}~\bibnamefont {Je{\v{z}}ek}},\ and\ \bibinfo
  {author} {\bibfnamefont {R.}~\bibnamefont {Filip}},\ }\href
  {https://journals.aps.org/prl/abstract/10.1103/PhysRevLett.123.043601}
  {\bibfield  {journal} {\bibinfo  {journal} {Phys. Rev. Lett.}\ }\textbf
  {\bibinfo {volume} {123}},\ \bibinfo {pages} {043601} (\bibinfo {year}
  {2019})}\BibitemShut {NoStop}%
\bibitem [{\citenamefont {Chabaud}\ \emph {et~al.}(2021)\citenamefont
  {Chabaud}, \citenamefont {Roeland}, \citenamefont {Walschaers}, \citenamefont
  {Grosshans}, \citenamefont {Parigi}, \citenamefont {Markham},\ and\
  \citenamefont {Treps}}]{chabaud2021certification}%
  \BibitemOpen
  \bibfield  {author} {\bibinfo {author} {\bibfnamefont {U.}~\bibnamefont
  {Chabaud}}, \bibinfo {author} {\bibfnamefont {G.}~\bibnamefont {Roeland}},
  \bibinfo {author} {\bibfnamefont {M.}~\bibnamefont {Walschaers}}, \bibinfo
  {author} {\bibfnamefont {F.}~\bibnamefont {Grosshans}}, \bibinfo {author}
  {\bibfnamefont {V.}~\bibnamefont {Parigi}}, \bibinfo {author} {\bibfnamefont
  {D.}~\bibnamefont {Markham}},\ and\ \bibinfo {author} {\bibfnamefont
  {N.}~\bibnamefont {Treps}},\ }\href@noop {} {\bibfield  {journal} {\bibinfo
  {journal} {PRX Quantum}\ }\textbf {\bibinfo {volume} {2}},\ \bibinfo {pages}
  {020333} (\bibinfo {year} {2021})}\BibitemShut {NoStop}%
\bibitem [{\citenamefont {Leibfried}\ \emph {et~al.}(1996)\citenamefont
  {Leibfried}, \citenamefont {Meekhof}, \citenamefont {King}, \citenamefont
  {Monroe}, \citenamefont {Itano},\ and\ \citenamefont
  {Wineland}}]{leibfried1996experimentalS}%
  \BibitemOpen
  \bibfield  {author} {\bibinfo {author} {\bibfnamefont {D.}~\bibnamefont
  {Leibfried}}, \bibinfo {author} {\bibfnamefont {D.}~\bibnamefont {Meekhof}},
  \bibinfo {author} {\bibfnamefont {B.}~\bibnamefont {King}}, \bibinfo {author}
  {\bibfnamefont {C.}~\bibnamefont {Monroe}}, \bibinfo {author} {\bibfnamefont
  {W.~M.}\ \bibnamefont {Itano}},\ and\ \bibinfo {author} {\bibfnamefont
  {D.~J.}\ \bibnamefont {Wineland}},\ }\href
  {https://journals.aps.org/prl/abstract/10.1103/PhysRevLett.77.4281}
  {\bibfield  {journal} {\bibinfo  {journal} {Phys. Rev. Lett.}\ }\textbf
  {\bibinfo {volume} {77}},\ \bibinfo {pages} {4281} (\bibinfo {year}
  {1996})}\BibitemShut {NoStop}%
\bibitem [{\citenamefont {Roos}\ \emph {et~al.}(1999)\citenamefont {Roos},
  \citenamefont {Zeiger}, \citenamefont {Rohde}, \citenamefont {N{\"a}gerl},
  \citenamefont {Eschner}, \citenamefont {Leibfried}, \citenamefont
  {Schmidt-Kaler},\ and\ \citenamefont {Blatt}}]{roos1999quantumS}%
  \BibitemOpen
  \bibfield  {author} {\bibinfo {author} {\bibfnamefont {C.}~\bibnamefont
  {Roos}}, \bibinfo {author} {\bibfnamefont {T.}~\bibnamefont {Zeiger}},
  \bibinfo {author} {\bibfnamefont {H.}~\bibnamefont {Rohde}}, \bibinfo
  {author} {\bibfnamefont {H.}~\bibnamefont {N{\"a}gerl}}, \bibinfo {author}
  {\bibfnamefont {J.}~\bibnamefont {Eschner}}, \bibinfo {author} {\bibfnamefont
  {D.}~\bibnamefont {Leibfried}}, \bibinfo {author} {\bibfnamefont
  {F.}~\bibnamefont {Schmidt-Kaler}},\ and\ \bibinfo {author} {\bibfnamefont
  {R.}~\bibnamefont {Blatt}},\ }\href
  {https://journals.aps.org/prl/abstract/10.1103/PhysRevLett.83.4713}
  {\bibfield  {journal} {\bibinfo  {journal} {Phys. Rev. Lett.}\ }\textbf
  {\bibinfo {volume} {83}},\ \bibinfo {pages} {4713} (\bibinfo {year}
  {1999})}\BibitemShut {NoStop}%
\bibitem [{\citenamefont {McCormick}\ \emph {et~al.}(2019)\citenamefont
  {McCormick}, \citenamefont {Keller}, \citenamefont {Burd}, \citenamefont
  {Wineland}, \citenamefont {Wilson},\ and\ \citenamefont
  {Leibfried}}]{mccormick2019quantumS}%
  \BibitemOpen
  \bibfield  {author} {\bibinfo {author} {\bibfnamefont {K.~C.}\ \bibnamefont
  {McCormick}}, \bibinfo {author} {\bibfnamefont {J.}~\bibnamefont {Keller}},
  \bibinfo {author} {\bibfnamefont {S.~C.}\ \bibnamefont {Burd}}, \bibinfo
  {author} {\bibfnamefont {D.~J.}\ \bibnamefont {Wineland}}, \bibinfo {author}
  {\bibfnamefont {A.~C.}\ \bibnamefont {Wilson}},\ and\ \bibinfo {author}
  {\bibfnamefont {D.}~\bibnamefont {Leibfried}},\ }\href
  {https://www.nature.com/articles/s41586-019-1421-y} {\bibfield  {journal}
  {\bibinfo  {journal} {Nature}\ }\textbf {\bibinfo {volume} {572}},\ \bibinfo
  {pages} {86} (\bibinfo {year} {2019})}\BibitemShut {NoStop}%
\bibitem [{\citenamefont {Cirac}\ \emph {et~al.}(1993)\citenamefont {Cirac},
  \citenamefont {Blatt}, \citenamefont {Parkins},\ and\ \citenamefont
  {Zoller}}]{cirac1993preparationS}%
  \BibitemOpen
  \bibfield  {author} {\bibinfo {author} {\bibfnamefont {J.~I.}\ \bibnamefont
  {Cirac}}, \bibinfo {author} {\bibfnamefont {R.}~\bibnamefont {Blatt}},
  \bibinfo {author} {\bibfnamefont {A.}~\bibnamefont {Parkins}},\ and\ \bibinfo
  {author} {\bibfnamefont {P.}~\bibnamefont {Zoller}},\ }\href
  {https://journals.aps.org/prl/abstract/10.1103/PhysRevLett.70.762} {\bibfield
   {journal} {\bibinfo  {journal} {Phys. Rev. Lett.}\ }\textbf {\bibinfo
  {volume} {70}},\ \bibinfo {pages} {762} (\bibinfo {year} {1993})}\BibitemShut
  {NoStop}%
\bibitem [{\citenamefont {Eschner}\ \emph {et~al.}(1995)\citenamefont
  {Eschner}, \citenamefont {Appasamy},\ and\ \citenamefont
  {Toschek}}]{eschner1995stochasticS}%
  \BibitemOpen
  \bibfield  {author} {\bibinfo {author} {\bibfnamefont {J.}~\bibnamefont
  {Eschner}}, \bibinfo {author} {\bibfnamefont {B.}~\bibnamefont {Appasamy}},\
  and\ \bibinfo {author} {\bibfnamefont {P.}~\bibnamefont {Toschek}},\ }\href
  {https://journals.aps.org/prl/abstract/10.1103/PhysRevLett.74.2435}
  {\bibfield  {journal} {\bibinfo  {journal} {Phys. Rev. Lett.}\ }\textbf
  {\bibinfo {volume} {74}},\ \bibinfo {pages} {2435} (\bibinfo {year}
  {1995})}\BibitemShut {NoStop}%
\bibitem [{\citenamefont {Cirac}\ \emph {et~al.}(1994)\citenamefont {Cirac},
  \citenamefont {Blatt},\ and\ \citenamefont
  {Zoller}}]{cirac1994nonclassicalS}%
  \BibitemOpen
  \bibfield  {author} {\bibinfo {author} {\bibfnamefont {J.~I.}\ \bibnamefont
  {Cirac}}, \bibinfo {author} {\bibfnamefont {R.}~\bibnamefont {Blatt}},\ and\
  \bibinfo {author} {\bibfnamefont {P.}~\bibnamefont {Zoller}},\ }\href
  {https://journals.aps.org/pra/abstract/10.1103/PhysRevA.49.R3174} {\bibfield
  {journal} {\bibinfo  {journal} {Phys. Rev. A}\ }\textbf {\bibinfo {volume}
  {49}},\ \bibinfo {pages} {R3174} (\bibinfo {year} {1994})}\BibitemShut
  {NoStop}%
\bibitem [{\citenamefont {Blatt}\ \emph {et~al.}(1995)\citenamefont {Blatt},
  \citenamefont {Cirac},\ and\ \citenamefont {Zoller}}]{blatt1995trappingS}%
  \BibitemOpen
  \bibfield  {author} {\bibinfo {author} {\bibfnamefont {R.}~\bibnamefont
  {Blatt}}, \bibinfo {author} {\bibfnamefont {J.~I.}\ \bibnamefont {Cirac}},\
  and\ \bibinfo {author} {\bibfnamefont {P.}~\bibnamefont {Zoller}},\ }\href
  {https://journals.aps.org/pra/abstract/10.1103/PhysRevA.52.518} {\bibfield
  {journal} {\bibinfo  {journal} {Phys. Rev. A}\ }\textbf {\bibinfo {volume}
  {52}},\ \bibinfo {pages} {518} (\bibinfo {year} {1995})}\BibitemShut
  {NoStop}%
\bibitem [{\citenamefont {Myatt}\ \emph {et~al.}(2000)\citenamefont {Myatt},
  \citenamefont {King}, \citenamefont {Turchette}, \citenamefont {Sackett},
  \citenamefont {Kielpinski}, \citenamefont {Itano}, \citenamefont {Monroe},\
  and\ \citenamefont {Wineland}}]{myatt2000decoherenceS}%
  \BibitemOpen
  \bibfield  {author} {\bibinfo {author} {\bibfnamefont {C.~J.}\ \bibnamefont
  {Myatt}}, \bibinfo {author} {\bibfnamefont {B.~E.}\ \bibnamefont {King}},
  \bibinfo {author} {\bibfnamefont {Q.~A.}\ \bibnamefont {Turchette}}, \bibinfo
  {author} {\bibfnamefont {C.~A.}\ \bibnamefont {Sackett}}, \bibinfo {author}
  {\bibfnamefont {D.}~\bibnamefont {Kielpinski}}, \bibinfo {author}
  {\bibfnamefont {W.~M.}\ \bibnamefont {Itano}}, \bibinfo {author}
  {\bibfnamefont {C.}~\bibnamefont {Monroe}},\ and\ \bibinfo {author}
  {\bibfnamefont {D.~J.}\ \bibnamefont {Wineland}},\ }\href
  {https://www.nature.com/articles/35002001} {\bibfield  {journal} {\bibinfo
  {journal} {Nature}\ }\textbf {\bibinfo {volume} {403}},\ \bibinfo {pages}
  {269} (\bibinfo {year} {2000})}\BibitemShut {NoStop}%
\bibitem [{\citenamefont {Podhora}\ \emph {et~al.}(2020)\citenamefont
  {Podhora}, \citenamefont {Pham}, \citenamefont {Le{\v{s}}und{\'a}k},
  \citenamefont {Ob{\v{s}}il}, \citenamefont {{\v{C}}{\'\i}{\v{z}}ek},
  \citenamefont {{\v{C}}{\'\i}p}, \citenamefont {Marek}, \citenamefont
  {Slodi{\v{c}}ka},\ and\ \citenamefont {Filip}}]{podhora2020unconditionalS}%
  \BibitemOpen
  \bibfield  {author} {\bibinfo {author} {\bibfnamefont {L.}~\bibnamefont
  {Podhora}}, \bibinfo {author} {\bibfnamefont {T.}~\bibnamefont {Pham}},
  \bibinfo {author} {\bibfnamefont {A.}~\bibnamefont {Le{\v{s}}und{\'a}k}},
  \bibinfo {author} {\bibfnamefont {P.}~\bibnamefont {Ob{\v{s}}il}}, \bibinfo
  {author} {\bibfnamefont {M.}~\bibnamefont {{\v{C}}{\'\i}{\v{z}}ek}}, \bibinfo
  {author} {\bibfnamefont {O.}~\bibnamefont {{\v{C}}{\'\i}p}}, \bibinfo
  {author} {\bibfnamefont {P.}~\bibnamefont {Marek}}, \bibinfo {author}
  {\bibfnamefont {L.}~\bibnamefont {Slodi{\v{c}}ka}},\ and\ \bibinfo {author}
  {\bibfnamefont {R.}~\bibnamefont {Filip}},\ }\href
  {https://onlinelibrary.wiley.com/doi/abs/10.1002/qute.202000012} {\bibfield
  {journal} {\bibinfo  {journal} {Adv. Quantum Technol.}\ }\textbf {\bibinfo
  {volume} {3}},\ \bibinfo {pages} {2000012} (\bibinfo {year}
  {2020})}\BibitemShut {NoStop}%
\bibitem [{\citenamefont {Wolf}\ \emph {et~al.}(2019)\citenamefont {Wolf},
  \citenamefont {Shi}, \citenamefont {Heip}, \citenamefont {Gessner},
  \citenamefont {Pezz{\`e}}, \citenamefont {Smerzi}, \citenamefont {Schulte},
  \citenamefont {Hammerer},\ and\ \citenamefont {Schmidt}}]{wolf2019motionalS}%
  \BibitemOpen
  \bibfield  {author} {\bibinfo {author} {\bibfnamefont {F.}~\bibnamefont
  {Wolf}}, \bibinfo {author} {\bibfnamefont {C.}~\bibnamefont {Shi}}, \bibinfo
  {author} {\bibfnamefont {J.~C.}\ \bibnamefont {Heip}}, \bibinfo {author}
  {\bibfnamefont {M.}~\bibnamefont {Gessner}}, \bibinfo {author} {\bibfnamefont
  {L.}~\bibnamefont {Pezz{\`e}}}, \bibinfo {author} {\bibfnamefont
  {A.}~\bibnamefont {Smerzi}}, \bibinfo {author} {\bibfnamefont
  {M.}~\bibnamefont {Schulte}}, \bibinfo {author} {\bibfnamefont
  {K.}~\bibnamefont {Hammerer}},\ and\ \bibinfo {author} {\bibfnamefont
  {P.~O.}\ \bibnamefont {Schmidt}},\ }\href
  {https://www.nature.com/articles/s41467-019-10576-4} {\bibfield  {journal}
  {\bibinfo  {journal} {Nat. Commun.}\ }\textbf {\bibinfo {volume} {10}},\
  \bibinfo {pages} {1} (\bibinfo {year} {2019})}\BibitemShut {NoStop}%
\bibitem [{\citenamefont {Um}\ \emph {et~al.}(2016)\citenamefont {Um},
  \citenamefont {Zhang}, \citenamefont {Lv}, \citenamefont {Lu}, \citenamefont
  {An}, \citenamefont {Zhang}, \citenamefont {Nha}, \citenamefont {Kim},\ and\
  \citenamefont {Kim}}]{um2016phononS}%
  \BibitemOpen
  \bibfield  {author} {\bibinfo {author} {\bibfnamefont {M.}~\bibnamefont
  {Um}}, \bibinfo {author} {\bibfnamefont {J.}~\bibnamefont {Zhang}}, \bibinfo
  {author} {\bibfnamefont {D.}~\bibnamefont {Lv}}, \bibinfo {author}
  {\bibfnamefont {Y.}~\bibnamefont {Lu}}, \bibinfo {author} {\bibfnamefont
  {S.}~\bibnamefont {An}}, \bibinfo {author} {\bibfnamefont {J.-N.}\
  \bibnamefont {Zhang}}, \bibinfo {author} {\bibfnamefont {H.}~\bibnamefont
  {Nha}}, \bibinfo {author} {\bibfnamefont {M.~S.}\ \bibnamefont {Kim}},\ and\
  \bibinfo {author} {\bibfnamefont {K.}~\bibnamefont {Kim}},\ }\href
  {https://www.nature.com/articles/ncomms11410} {\bibfield  {journal} {\bibinfo
   {journal} {Nat. Commun.}\ }\textbf {\bibinfo {volume} {7}},\ \bibinfo
  {pages} {11410} (\bibinfo {year} {2016})}\BibitemShut {NoStop}%
\bibitem [{\citenamefont {Kienzler}\ \emph {et~al.}(2015)\citenamefont
  {Kienzler}, \citenamefont {Lo}, \citenamefont {Keitch}, \citenamefont
  {De~Clercq}, \citenamefont {Leupold}, \citenamefont {Lindenfelser},
  \citenamefont {Marinelli}, \citenamefont {Negnevitsky},\ and\ \citenamefont
  {Home}}]{kienzler2015quantumS}%
  \BibitemOpen
  \bibfield  {author} {\bibinfo {author} {\bibfnamefont {D.}~\bibnamefont
  {Kienzler}}, \bibinfo {author} {\bibfnamefont {H.-Y.}\ \bibnamefont {Lo}},
  \bibinfo {author} {\bibfnamefont {B.}~\bibnamefont {Keitch}}, \bibinfo
  {author} {\bibfnamefont {L.}~\bibnamefont {De~Clercq}}, \bibinfo {author}
  {\bibfnamefont {F.}~\bibnamefont {Leupold}}, \bibinfo {author} {\bibfnamefont
  {F.}~\bibnamefont {Lindenfelser}}, \bibinfo {author} {\bibfnamefont
  {M.}~\bibnamefont {Marinelli}}, \bibinfo {author} {\bibfnamefont
  {V.}~\bibnamefont {Negnevitsky}},\ and\ \bibinfo {author} {\bibfnamefont
  {J.~P.}\ \bibnamefont {Home}},\ }\href
  {https://www.science.org/doi/abs/10.1126/science.1261033} {\bibfield
  {journal} {\bibinfo  {journal} {Science}\ }\textbf {\bibinfo {volume}
  {347}},\ \bibinfo {pages} {53} (\bibinfo {year} {2015})}\BibitemShut
  {NoStop}%
\bibitem [{\citenamefont {Meekhof}\ \emph {et~al.}(1996)\citenamefont
  {Meekhof}, \citenamefont {Monroe}, \citenamefont {King}, \citenamefont
  {Itano},\ and\ \citenamefont {Wineland}}]{meekhof1996generationS}%
  \BibitemOpen
  \bibfield  {author} {\bibinfo {author} {\bibfnamefont {D.}~\bibnamefont
  {Meekhof}}, \bibinfo {author} {\bibfnamefont {C.}~\bibnamefont {Monroe}},
  \bibinfo {author} {\bibfnamefont {B.}~\bibnamefont {King}}, \bibinfo {author}
  {\bibfnamefont {W.~M.}\ \bibnamefont {Itano}},\ and\ \bibinfo {author}
  {\bibfnamefont {D.~J.}\ \bibnamefont {Wineland}},\ }\href
  {https://journals.aps.org/prl/abstract/10.1103/PhysRevLett.76.1796}
  {\bibfield  {journal} {\bibinfo  {journal} {Phys. Rev. Lett.}\ }\textbf
  {\bibinfo {volume} {76}},\ \bibinfo {pages} {1796} (\bibinfo {year}
  {1996})}\BibitemShut {NoStop}%
\bibitem [{\citenamefont {Schneider}\ and\ \citenamefont
  {Milburn}(1998)}]{schneider1998decoherenceS}%
  \BibitemOpen
  \bibfield  {author} {\bibinfo {author} {\bibfnamefont {S.}~\bibnamefont
  {Schneider}}\ and\ \bibinfo {author} {\bibfnamefont {G.~J.}\ \bibnamefont
  {Milburn}},\ }\href
  {https://journals.aps.org/pra/abstract/10.1103/PhysRevA.57.3748} {\bibfield
  {journal} {\bibinfo  {journal} {Phys. Rev. A}\ }\textbf {\bibinfo {volume}
  {57}},\ \bibinfo {pages} {3748} (\bibinfo {year} {1998})}\BibitemShut
  {NoStop}%
\bibitem [{\citenamefont {Murao}\ and\ \citenamefont
  {Knight}(1998)}]{murao1998decoherence}%
  \BibitemOpen
  \bibfield  {author} {\bibinfo {author} {\bibfnamefont {M.}~\bibnamefont
  {Murao}}\ and\ \bibinfo {author} {\bibfnamefont {P.}~\bibnamefont {Knight}},\
  }\href {https://journals.aps.org/pra/abstract/10.1103/PhysRevA.58.663}
  {\bibfield  {journal} {\bibinfo  {journal} {Phys. Rev. A}\ }\textbf {\bibinfo
  {volume} {58}},\ \bibinfo {pages} {663} (\bibinfo {year} {1998})}\BibitemShut
  {NoStop}%
\bibitem [{\citenamefont {Bonifacio}\ \emph {et~al.}(2000)\citenamefont
  {Bonifacio}, \citenamefont {Olivares}, \citenamefont {Tombesi},\ and\
  \citenamefont {Vitali}}]{bonifacio2000model}%
  \BibitemOpen
  \bibfield  {author} {\bibinfo {author} {\bibfnamefont {R.}~\bibnamefont
  {Bonifacio}}, \bibinfo {author} {\bibfnamefont {S.}~\bibnamefont {Olivares}},
  \bibinfo {author} {\bibfnamefont {P.}~\bibnamefont {Tombesi}},\ and\ \bibinfo
  {author} {\bibfnamefont {D.}~\bibnamefont {Vitali}},\ }\href
  {https://journals.aps.org/pra/abstract/10.1103/PhysRevA.61.053802} {\bibfield
   {journal} {\bibinfo  {journal} {Phys. Rev. A}\ }\textbf {\bibinfo {volume}
  {61}},\ \bibinfo {pages} {053802} (\bibinfo {year} {2000})}\BibitemShut
  {NoStop}%
\bibitem [{\citenamefont {Budini}\ \emph {et~al.}(2002)\citenamefont {Budini},
  \citenamefont {de~Matos~Filho},\ and\ \citenamefont
  {Zagury}}]{budini2002localization}%
  \BibitemOpen
  \bibfield  {author} {\bibinfo {author} {\bibfnamefont {A.}~\bibnamefont
  {Budini}}, \bibinfo {author} {\bibfnamefont {R.}~\bibnamefont
  {de~Matos~Filho}},\ and\ \bibinfo {author} {\bibfnamefont {N.}~\bibnamefont
  {Zagury}},\ }\href
  {https://journals.aps.org/pra/abstract/10.1103/PhysRevA.65.041402} {\bibfield
   {journal} {\bibinfo  {journal} {Phys. Rev. A}\ }\textbf {\bibinfo {volume}
  {65}},\ \bibinfo {pages} {041402} (\bibinfo {year} {2002})}\BibitemShut
  {NoStop}%
\bibitem [{\citenamefont {Leibfried}\ \emph {et~al.}(2003)\citenamefont
  {Leibfried}, \citenamefont {Blatt}, \citenamefont {Monroe},\ and\
  \citenamefont {Wineland}}]{leibfried2003quantumS}%
  \BibitemOpen
  \bibfield  {author} {\bibinfo {author} {\bibfnamefont {D.}~\bibnamefont
  {Leibfried}}, \bibinfo {author} {\bibfnamefont {R.}~\bibnamefont {Blatt}},
  \bibinfo {author} {\bibfnamefont {C.}~\bibnamefont {Monroe}},\ and\ \bibinfo
  {author} {\bibfnamefont {D.}~\bibnamefont {Wineland}},\ }\href
  {https://journals.aps.org/rmp/abstract/10.1103/RevModPhys.75.281} {\bibfield
  {journal} {\bibinfo  {journal} {Rev. Mod. Phys.}\ }\textbf {\bibinfo {volume}
  {75}},\ \bibinfo {pages} {281} (\bibinfo {year} {2003})}\BibitemShut
  {NoStop}%
\bibitem [{\citenamefont {Brownnutt}\ \emph {et~al.}(2015)\citenamefont
  {Brownnutt}, \citenamefont {Kumph}, \citenamefont {Rabl},\ and\ \citenamefont
  {Blatt}}]{brownnutt2015ionS}%
  \BibitemOpen
  \bibfield  {author} {\bibinfo {author} {\bibfnamefont {M.}~\bibnamefont
  {Brownnutt}}, \bibinfo {author} {\bibfnamefont {M.}~\bibnamefont {Kumph}},
  \bibinfo {author} {\bibfnamefont {P.}~\bibnamefont {Rabl}},\ and\ \bibinfo
  {author} {\bibfnamefont {R.}~\bibnamefont {Blatt}},\ }\href
  {https://journals.aps.org/rmp/pdf/10.1103/RevModPhys.87.1419} {\bibfield
  {journal} {\bibinfo  {journal} {Rev. Mod. Phys.}\ }\textbf {\bibinfo {volume}
  {87}},\ \bibinfo {pages} {1419} (\bibinfo {year} {2015})}\BibitemShut
  {NoStop}%
\bibitem [{\citenamefont {Aspelmeyer}\ \emph {et~al.}(2014)\citenamefont
  {Aspelmeyer}, \citenamefont {Kippenberg},\ and\ \citenamefont
  {Marquardt}}]{aspelmeyer2014cavityS}%
  \BibitemOpen
  \bibfield  {author} {\bibinfo {author} {\bibfnamefont {M.}~\bibnamefont
  {Aspelmeyer}}, \bibinfo {author} {\bibfnamefont {T.~J.}\ \bibnamefont
  {Kippenberg}},\ and\ \bibinfo {author} {\bibfnamefont {F.}~\bibnamefont
  {Marquardt}},\ }\href
  {https://journals.aps.org/rmp/abstract/10.1103/RevModPhys.86.1391} {\bibfield
   {journal} {\bibinfo  {journal} {Rev. Mod. Phys.}\ }\textbf {\bibinfo
  {volume} {86}},\ \bibinfo {pages} {1391} (\bibinfo {year}
  {2014})}\BibitemShut {NoStop}%
\bibitem [{\citenamefont {Eschner}\ \emph {et~al.}(2003)\citenamefont
  {Eschner}, \citenamefont {Morigi}, \citenamefont {Schmidt-Kaler},\ and\
  \citenamefont {Blatt}}]{eschner2003laserS}%
  \BibitemOpen
  \bibfield  {author} {\bibinfo {author} {\bibfnamefont {J.}~\bibnamefont
  {Eschner}}, \bibinfo {author} {\bibfnamefont {G.}~\bibnamefont {Morigi}},
  \bibinfo {author} {\bibfnamefont {F.}~\bibnamefont {Schmidt-Kaler}},\ and\
  \bibinfo {author} {\bibfnamefont {R.}~\bibnamefont {Blatt}},\ }\href
  {https://www.osapublishing.org/josab/abstract.cfm?uri=JOSAB-20-5-1003}
  {\bibfield  {journal} {\bibinfo  {journal} {J. Opt. Soc. Am. B}\ }\textbf
  {\bibinfo {volume} {20}},\ \bibinfo {pages} {1003} (\bibinfo {year}
  {2003})}\BibitemShut {NoStop}%
\end{thebibliography}
\end{document}